\documentclass[aps,prc,twocolumn,superscriptaddress,showpacs]{revtex4-1}

\usepackage[]{graphicx}
\usepackage{epstopdf}
\usepackage[]{epsfig}
\usepackage{longtable}
\usepackage{hyperref}
\hypersetup{letterpaper, colorlinks=true, urlcolor=blue, citecolor=blue, linkcolor=blue, breaklinks=true}
\usepackage{breakurl}
\usepackage{booktabs}
\usepackage{subfigure}
\usepackage{tabularx}
\usepackage{braket}
\usepackage{float}

\begin{document}

\title{Understanding the Low-Energy Enhancement of the $\gamma$-ray Strength Function of $^{56}$Fe.}


\author{M.D.~Jones}
    \affiliation{Nuclear Science Division, Lawrence Berkeley National Laboratory, Berkeley, CA 94720, USA}
\author{A.O.~Macchiavelli}
   \affiliation{Nuclear Science Division, Lawrence Berkeley National Laboratory, Berkeley, CA 94720, USA}
\author{M.~Wiedeking}
    \affiliation{iThemba LABS, P.O. Box 722, Somerset West 7129, South Africa}
\author{L.A.~Bernstein}
   \affiliation{Nuclear Science Division, Lawrence Berkeley National Laboratory, Berkeley, CA 94720, USA}
\author{H.L.~Crawford}
   \affiliation{Nuclear Science Division, Lawrence Berkeley National Laboratory, Berkeley, CA 94720, USA}
\author{C.M.~Campbell}
   \affiliation{Nuclear Science Division, Lawrence Berkeley National Laboratory, Berkeley, CA 94720, USA}
\author{R.M.~Clark}
  \affiliation{Nuclear Science Division, Lawrence Berkeley National Laboratory, Berkeley, CA 94720, USA}
\author{M.~Cromaz}
   \affiliation{Nuclear Science Division, Lawrence Berkeley National Laboratory, Berkeley, CA 94720, USA}
\author{P.~Fallon}
  \affiliation{Nuclear Science Division, Lawrence Berkeley National Laboratory, Berkeley, CA 94720, USA}
\author{I.Y.~Lee}
   \affiliation{Nuclear Science Division, Lawrence Berkeley National Laboratory, Berkeley, CA 94720, USA}
\author{M.~Salathe}
   \affiliation{Nuclear Science Division, Lawrence Berkeley National Laboratory, Berkeley, CA 94720, USA}
\author{A.~Wiens}
   \altaffiliation[Present Address: ]{Comprehensive Nuclear-Test-Ban Treaty Organization (CTBTO), Vienna, Austria}
   \affiliation{Nuclear Science Division, Lawrence Berkeley National Laboratory, Berkeley, CA 94720, USA}
 \author{A.D.~Ayangeakaa}
   \altaffiliation[Present Address: ]{Department of Physics, United States Naval Academy, Annapolis, MD 21402, USA}
   \affiliation{Physics Division, Argonne National Laboratory, Argonne, Illinois 60439, USA}
\author{D.L. Bleuel}
   \affiliation{Physical and Life Sciences Directorate, Lawrence Livermore National Laboratory, Livermore, California 94551, USA} 
\author{S.~Bottoni}
   \altaffiliation[Present Address: ]{Universit\`a degli Studi di Milano and INFN, Via Celoria 16, I-20133 Milano, Italy}
    \affiliation{Physics Division, Argonne National Laboratory, Argonne, Illinois 60439, USA}
\author{M.P.~Carpenter}
   \affiliation{Physics Division, Argonne National Laboratory, Argonne, Illinois 60439, USA}
\author{H.M.~Davids}
	\altaffiliation[Present Address: ]{GSI,  Planckstra\ss e 1, D-64291, Darmstadt, Germany}
   \affiliation{Physics Division, Argonne National Laboratory, Argonne, Illinois 60439, USA}  
\author{J.~Elson}
   \affiliation{Department of Chemistry, Washington University, St. Louis, Missouri 63130, USA}
 \author{A. ~G\"orgen}
   \affiliation{Department of Physics, University of Oslo, N-0316 Oslo, Norway}  
  \author{M.~Guttormsen}
   \affiliation{Department of Physics, University of Oslo, N-0316 Oslo, Norway} 
\author{R.V.F.~Janssens}
   \altaffiliation[Present Address: ]{Department of Physics and Astronomy, University of North Carolina at Chapel Hill, Chapel Hill, NC 27559-3255, USA and TUNL, Duke University, Durham, NC, 27708-0308, USA}
   \affiliation{Physics Division, Argonne National Laboratory, Argonne, Illinois 60439, USA}  
 \author{J.E.~Kinnison }
   \affiliation{Department of Chemistry, Washington University, St. Louis, Missouri 63130, USA}
   \author{L.~Kirsch}
   \affiliation{Department of Nuclear Engineering, University of California, Berkeley, California 94720, USA}
\author{A.C.~Larsen}
   \affiliation{Department of Physics, University of Oslo, N-0316 Oslo, Norway}
\author{T.~Lauritsen}
   \affiliation{Physics Division, Argonne National Laboratory, Argonne, Illinois 60439, USA}
\author{W.~Reviol}
   \affiliation{Department of Chemistry, Washington University, St. Louis, Missouri 63130, USA}
\author{D.G.~Sarantites}
   \affiliation{Department of Chemistry, Washington University, St. Louis, Missouri 63130, USA}
\author{S.~Siem}
   \affiliation{Department of Physics, University of Oslo, N-0316 Oslo, Norway}

\author{A.V.~Voinov}
   \affiliation{Department of Physics and Astronomy, Ohio University, Athens, Ohio 45701, USA}
   
   \author{S.~Zhu}
   \affiliation{Physics Division, Argonne National Laboratory, Argonne, Illinois 60439, USA}







\begin{abstract}
    
A model-independent technique was used to determine the $\gamma$-ray Strength Function ($\gamma$SF) of $^{56}$Fe down to $\gamma$-ray energies less than 1 MeV for the first time with GRETINA using the $(p,p')$ reaction at 16 MeV.  No difference was observed in the energy dependence of the $\gamma$SF built on $2^{+}$ and $4^{+}$ final states, supporting the Brink hypothesis.   In addition, angular distribution and polarization measurements were performed. The angular distributions are consistent with dipole radiation.  The polarization results show a small bias towards magnetic character in the region of the enhancement.  


\end{abstract}


\maketitle

\section{Introduction}

The $\gamma$-ray Strength Function ($\gamma$SF) describes the statistical $\gamma$-ray decay properties of nucleonic systems at high excitation energy and level density \cite{Bar73}, and provides insight into the average reduced $\gamma$-ray transition probability for a given transition energy ($E_{\gamma}$) and multipolarity. The $\gamma$SF is dominated by the giant electric dipole resonance (GEDR) \cite{Har01}, a collective motion of neutrons against protons, at energies above the neutron threshold. The low-energy tail of the GEDR often exhibits other structural features which shed light on the underlying excitations modes of the nucleus e.g. the $E1$ pygmy \cite{Bra15, SAVRAN2013210}, $M1$ scissors \cite{Sch06}, or $M1$ spin-flip \cite{Neu10} resonances.

Statistical properties, such as the $\gamma$SF and Nuclear Level Density (NLD), are instrumental in describing photo-nuclear processes and neutron capture reaction rates \cite{GORIELY199810} as they are critical input parameters to the Hauser-Feshbach model for capture cross section calculations \cite{Hau52}.  The $\gamma$SF strongly affects capture cross sections and has the potential for far reaching impact on nucleosynthesis processes \cite{Arn03,Arn07}, nuclear waste transmutation \cite{Col10}, and nuclear fuel cycles \cite{Report06}. For instance, it has been shown that the presence of a Pygmy resonance \cite{GORIELY199810} or an enhanced low-energy $\gamma$-ray decay probability \cite{PhysRevC.82.014318} can lead to order of magnitude deviations on the capture cross sections for nuclei that undergo the rapid neutron-capture process \cite{RevModPhys.29.547}. The $\gamma$SF and NLD have been shown to reliably reproduce results from directly measured $(n,\gamma)$ \cite{Lap16, Khe17} and $(p,\gamma)$ \cite{Lar16} cross sections. Direct measurements are limited to reasonably long-lived targets and hence statistical properties will play an increasingly important role in determining many astrophysically relevant cross sections.  Experimental efforts already focus on new techniques, utilizing beta decay \cite{Spy14, Lid16} and surrogate reactions \cite{PhysRevLett.109.172501}, with the goal to obtain $(n,\gamma)$ cross sections for nuclei far from stability. 

\begin{figure*}[ht!]
\includegraphics[width=1.0\textwidth]{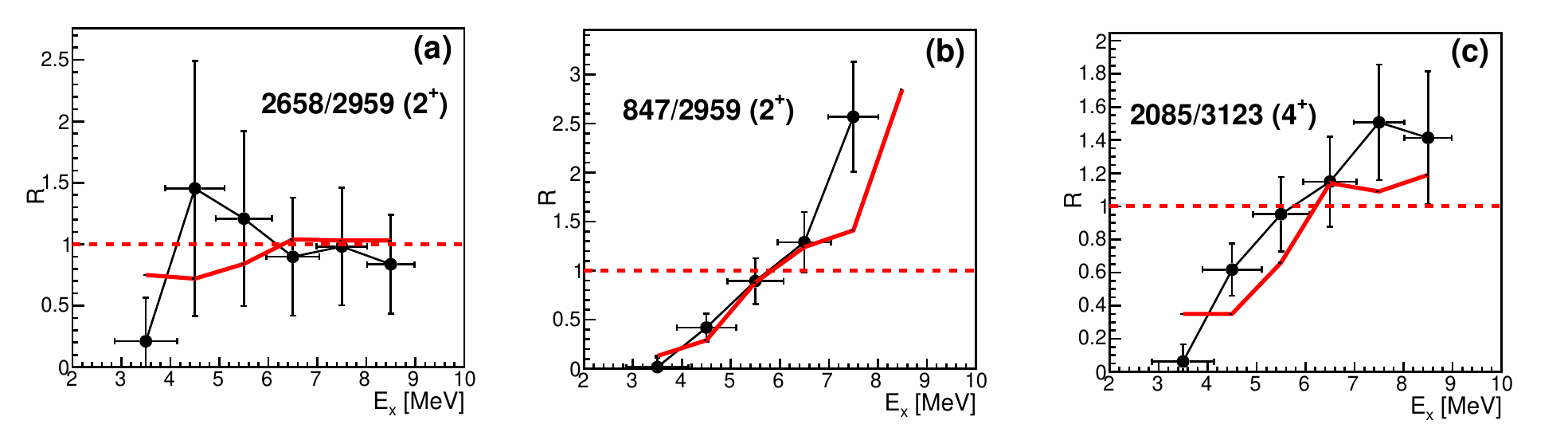}
\caption{The ratio $R = f(E_{i})/f(E_{j})$ as a function of excitation energy $E_{x}$, for several pairs of $2^{+}$ and $4^{+}$ states in $^{56}$Fe. The red-dashed line denotes $R=1$, while the solid red-curve shows the expected ratios from the polynomial fit in Figure \ref{fig:psf}. The final states used to construct the ratio are denoted by the fraction in the upper left or right with energies in keV.} 
\label{fig:ratio}
\end{figure*}

A low-energy enhancement ($E_{\gamma}< 4$~MeV) in the $\gamma$SF of $^{56}$Fe was discovered in 2004 \cite{PhysRevLett.93.142504}. This feature has been confirmed in a number of other light- and medium-mass nuclei, from $^{44}$Sc \cite{Lar07} to Cd isotopes \cite{Lar13} using the Oslo method \cite{GUTTORMSEN1987518,GUTTORMSEN1996371}. Recently, the enhancement has also been reported in the heavier rare-earth \cite{PhysRevC.93.034303} and lanthanide regions \cite{Khe15}. Furthermore, the existence of the enhancement was independently confirmed using the Ratio Method in $^{95}$Mo \cite{PhysRevLett.108.162503}, and these observations have spurred intense theoretical investigations. Shell-model calculations in the A$\sim$90 region have suggested the enhancement to be due to a large $B(M1)$ strength for low-energy $\gamma$-rays which is caused by orbital angular momentum recoupling of high-j orbits \cite{PhysRevLett.111.232504}. Calculations in $^{56}$Fe \cite{PhysRevLett.113.252502} and in $^{44}$Sc \cite{Sie17} have further revealed that $M1$ transitions, responsible for the enhancement, originate from 0$\hbar\omega$ states. However, other theoretical approaches propose an $E1$ strength to be responsible for the enhancement \cite{PhysRevC.88.031302}. 

Despite its broad impact, very little is known about the excitation mode responsible for the emergence of the low-energy enhancement. While recent measurements have demonstrated that the enhancement is dominated by dipole radiation \cite{PhysRevLett.111.242504, Lar17, PhysRevC.93.034303}, its electric or magnetic character remains an open question. A study on the total conversion coefficient of the $\gamma$-ray continuum in $^{163,163}$Yb formed in the $^{150}$Nd($^{20}$Ne, $xn\gamma$) reaction  indicated considerable contributions from $M1$ radiation near $E_{\gamma} \sim 500~$keV. In addition, studies of capture reactions in $^{59}$Co \cite{PhysRevC.81.024319} and  $^{144}$Nd \cite{PhysRevC.92.064308} have infered an $M1$ nature. A direct measurement of the polariation is the crucial missing piece of information which would constrain models and provide for a full understanding of the low-energy enhancement. In this article, we report the first polarization measurement of photons originating from the low-energy enhancement in the $\gamma$SF.

\section{Experimental Method}
The experiment was performed at the ATLAS facility at Argonne National Laboratory where a 16~MeV proton beam impinged upon a 1~mg/cm$^{2}$ 99.7$\%$ isotopically-enriched $^{56}$Fe target with an intensity of 0.50~pnA.  The target was surrounded by GRETINA \cite{PASCHALIS201344} (Gamma-Ray Energy Tracking In-beam Nuclear Array), and the Washington University Phoswich Wall \cite{Sarantites201542}. Eight GRETINA modules were positioned at a nominal distance of 18 cm around the target with one at 59$^{\circ}$, three at  90$^{\circ}$, and two at 121$^{\circ}$ and 147$^{\circ}$.  The singles photopeak efficiency at 1.33 MeV was $4.8\%$.
The hardware event trigger required that a Phoswich Wall element fired in coincidence with GRETINA within a 500 ns gate. A narrow coincidence gate of 10 ns was applied in the offline analysis.

The Phoswich Wall consists of four 64-fold-pixelated photomultiplier tubes with 2.2~mm thick CsI(Tl) and 12~$\mu$m thick fast-plastic scintillation detectors, having a total of 256 elements. To protect the detectors from the unreacted beam, their range of laboratory angles was chosen to be  32$^{\circ}$ $< \theta_{lab} < 75^{\circ}$, and the scintillators were covered with 100~$\mu$m thick Sn absorbers. The latter were supported by masks that slightly reduced the area of each pixel. The combined energy (CsI(Tl)) and energy loss (fast-plastic) information was used for particle identification, and the detector geometry allowed the excitation energy $E_{x}$ of the recoiling $^{56}$Fe nuclei to be deduced from the kinematics of the scattered protons.

\begin{figure}[t!]
\centering
\includegraphics[trim=5 0 30 30, clip, width=1.0\linewidth]{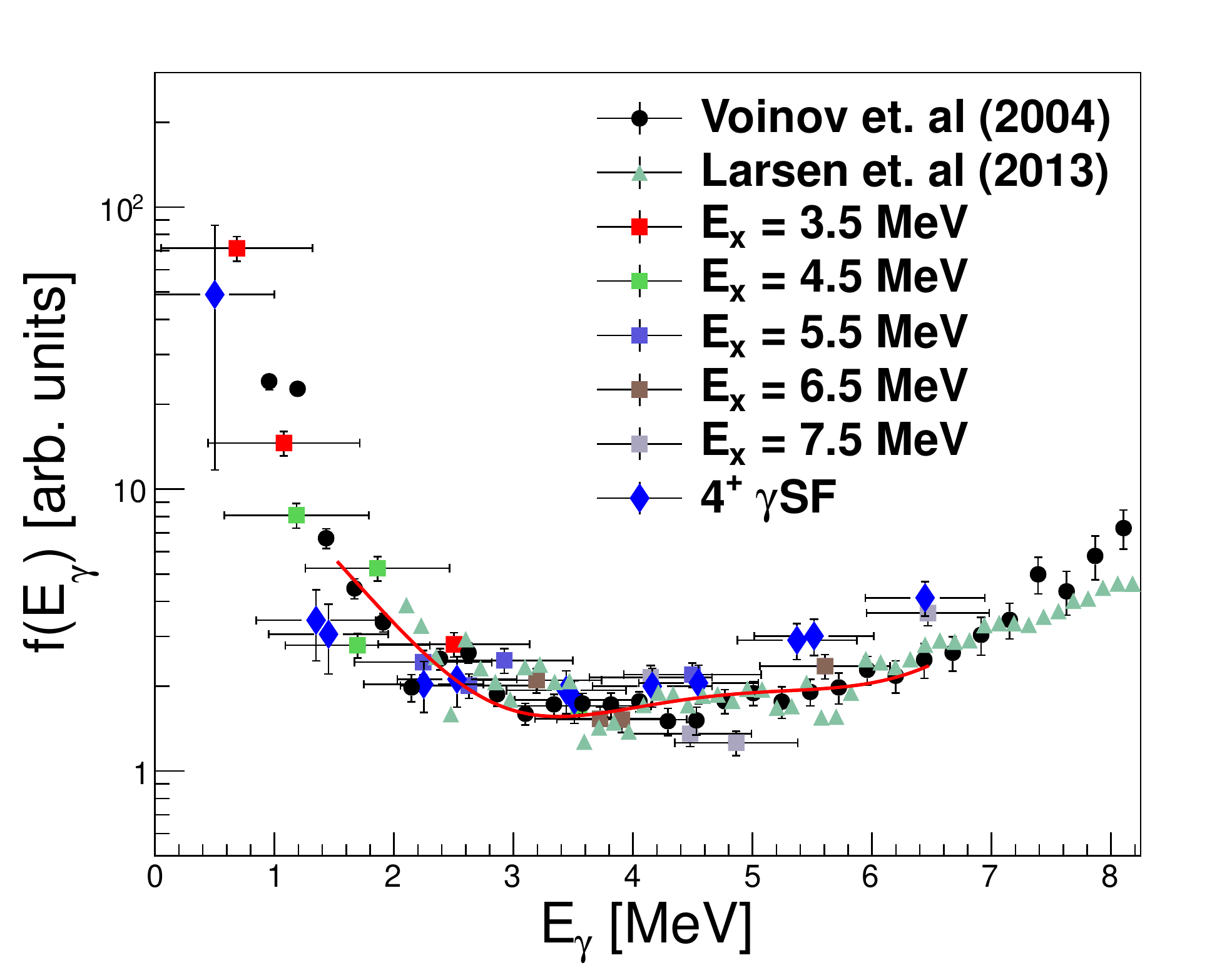}
\caption{(Color online) Gamma strength function for $^{56}$Fe from the present work (colored-squares) in comparison to previous measurements, Ref. \cite{PhysRevLett.93.142504} (black-circles) and Ref. \cite{PhysRevLett.111.242504} (green-triangles). 
The polynomial fit to Voinov \emph{et al.} \cite{PhysRevLett.93.142504} is shown by the red-solid curve.}
\label{fig:psf}
\end{figure}


The experiment was designed to measure statistical feeding from the quasicontinuum in $^{56}$Fe to specific low-lying states.  This was achieved with particle-$\gamma$-$\gamma$ triple coincidence events. Excited states in the quasicontinuum  were populated by the ($p,p^{`}$) reaction, and $\gamma$-rays originating from the quasicontinuum were identified by gating on the entrance excitation energy $E_{x}$ and on specific low-lying discrete transitions.

Any photon in coincidence with a proton and a discrete transition satisfying the energy difference $E_{\gamma} = E_{x} - E_{L} \pm \delta E$, where $E_{L}$ is the energy of the low-lying state and $\delta E$ the resolution in $E_{x}$, has an unambiguous origin and destination. Thus the intensity of single-step feeding to individual levels can be extracted on an event-by-event basis.

The $\gamma$SF was extracted via the Ratio Method \cite{PhysRevLett.108.162503}, briefly summarized here. The $\gamma$SF can be written as \cite{Bar73}:

\[ f(E_{\gamma}) = \frac{\braket{\Gamma_{J^{\pi}}(E_{x}, E_{\gamma})}\rho_{J^{\pi}}(E_{x})}{E_{\gamma}^{2\lambda +1}}   \]

\noindent where $\braket{\Gamma_{J^{\pi}}(E_{x}, E_{\gamma})}$ is the average radiative width, $\rho_{J^{\pi}}(E_{x})$ the level density, and $E_{\gamma}$ and $\lambda$ the energy and multipolarity of the first $\gamma$-ray emitted in the de-excitation of $^{56}$Fe. Assuming dipole radiation dominates, the intensity of $\gamma$-rays populating a specific low-lying state can be expressed in terms of the $\gamma$SF:

\[N_{i} \propto f(E_{i})E_{i}^{3}\sum_{J^{\pi}}  \sigma_{J^{\pi}} (E_{x}) \]

\noindent where the term $\sum_{J^{\pi}} \sigma_{J^{\pi}}(E_{x})$ denotes the cross section for populating a specific level in the reaction.

Let $E_{i}$ and $E_{j}$ denote two primary $\gamma$-rays feeding separate states of the same spin-parity. For a given excitation energy, the ratio of feeding from states in the quasicontinuum to a pair of low-lying states is proportional to the ratio of the strength function evaluated at $E_{i}$ and $E_{j}$:

\[ R = \left( \frac{N_{i}}{N_{j}} \right) \left( \frac{E_{j}}{E_{i}}\right)^{3} = \frac{f(E_{i})}{f(E_{j})} \]

By forming the ratio $R$ for discrete states of the same spin-parity, the dependence on the density of states in addition to other experimental systematic errors are removed, and the shape of the $\gamma$SF can be deduced.


\section{Analysis and Discussion}

The data were sorted on a calorimeter condition requiring that the total energy measured in GRETINA was equal to the measured excitation energy of $^{56}$Fe.  A total of 6 states had sufficient statistics to obtain ratios: four 2$^{+}$ states (847, 2658, 2959, 3369~keV) and two $4^{+}$ states (2085, 3123~keV). These levels and their branching ratios have been identified in previous experiments \cite{JUNDE20111513}.

Three of the seven ratios are shown in Figure \ref{fig:ratio} for two pairs of $2^{+}$ states (Fig.~\ref{fig:ratio} (a,b)) and the single pair of $4^{+}$ states (Fig.~\ref{fig:ratio}(c)). The uncertainty in the ratios is a combination of statistics and the error propagated from the resolution of the Phoswich Wall. 
The red curve represents the theoretical ratios obtained from a polynomial fit of the strength functions reported in previous measurements \cite{PhysRevLett.93.142504, PhysRevLett.111.242504}, and corresponds to the red curve in Figure \ref{fig:psf}. Good agreement with the Oslo method is observed \cite{GUTTORMSEN1996371, GUTTORMSEN1987518, PhysRevLett.111.242504}.

For a given excitation energy, the energy difference between the pair of discrete states is equal to the distance 
between the two points on $f(E_{\gamma})$ whose ratio is being measured. 
When the pair of discrete states is sufficiently close such that the strength function does not change quickly over their energy difference, it is expected that $R=1$ for all $E_{x}$. This is what is observed in Figure \ref{fig:ratio} (a) where the two states are separated by only 300~keV. The ratio is consistent with unity for $E_{x} \geq 4.5$~MeV. However, for the lowest point at $E_{x} = 3.5$~MeV, it dips suddenly. This implies that the strength function is increasing rapidly between $E_{\gamma} \sim 800$~keV and $E_{\gamma} \sim 500$~keV, which is consistent with a large low-energy enhancement. 
Figures \ref{fig:ratio} (b,c), show a general trend of $R < 1$ at low $E_{x}$, and hence low $E_{\gamma}$, that monotonically increases past $R=1$. This is indicative of a local minimum in the strength function. 

The ratios can be translated to individual points on the strength function $f(E_{\gamma})$, however the normalization between excitation energy bins is unconstrained. 
 For the purpose of comparison to previous data, the normalizations are minimized to a polynomial fit of the Oslo measurements \cite{PhysRevLett.93.142504, PhysRevLett.111.242504} between 1.5 and 4.5 MeV, shown in Figure \ref{fig:psf}. The low-energy enhancement is evident and appears to continue to increase below 1 MeV.


In addition to the $2^{+}$ states, a pair of $4^{+}$ states had sufficient statistics to form ratios. They are given the same normalization for comparison, shown in Figure \ref{fig:psf}. The strength function obtained from $4^{+}$ final states agrees with that obtained from $2^{+}$ final states, which is consistent with the Brink hypothesis. 

\subsection{Angular Distributions and Polarization}

Moving beyond the shape of the $\gamma$SF, the tracking capabilities of GRETINA allow one to obtain angular distribution and polarization information on the $\gamma$-rays in the region of the low-energy enhancement. The angular distributions are given by the intensity as a function of the lab angle, $\theta$ \cite{DERMATEOSIAN1974391}: 

\[ W(\theta) = 1 + a_{2}P_{2}(cos\theta) + a_{4}P_{4}(cos\theta), \]

\noindent where $P_{l}$ are the Legendre polynomials of degree $\ell$. The normalized angular-distribution coefficients are given by $a_{l} = Q_{l} \alpha_{l} A^{max}_{l}$, where $Q_{l}$ is the geometric attenuation of GRETINA, $A_{l}^{max}$ the coefficients for maximum alignment, and $\alpha_{l}$ the attenuation due to partial alignment. 

\begin{figure}[t!]
\includegraphics[trim=10 0 0 0 , clip, width=1.0\linewidth]{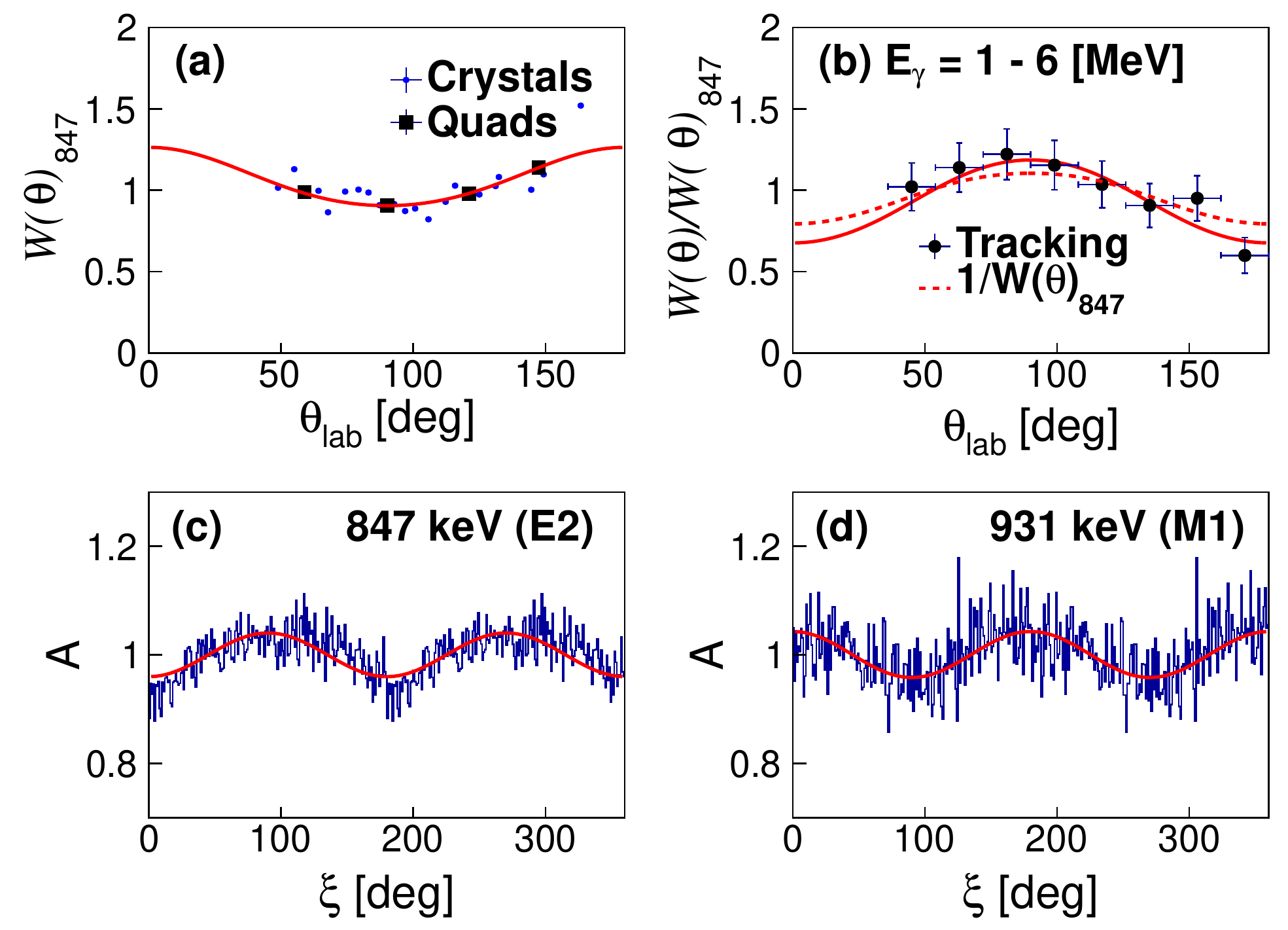}
\caption{(Color online) (a) Angular distribution for the 847~keV $2^{+}$ state in $^{56}$Fe. (b) The relative angular distribution for quasicontinuum $\gamma$-rays between 1.0 and 6.0 MeV. The dashed line shows $1/W(\theta)_{847}$. The bottom panels show the polarization asymmetry $A$ for an electric (c) and magnetic (d) transition in $^{56}$Fe and $^{55}$Fe respectively, with fits in solid red.} 
\label{fig:concept}
\end{figure}

The angular distributions for the quasicontinuum can be found in Figure \ref{fig:concept}(b), where a cut is made between $E_{\gamma} = 1 - 6$ MeV.  In order to remove systematics resulting from the triple-coincidence gate,  
the angular distributions are taken relative to the first-excited state in $^{56}$Fe (847 keV) with the same gating conditions. In order to extract the $a_{2}$ and $a_{4}$ coefficients, the ratio must be fit.

Using the measured values for the first-excited state of $^{56}$Fe (Fig. \ref{fig:concept}(a)), of $a_{2} = 0.22(5)$ and $a_{4} = 0.02(5)$, the extracted $a_{2}$ and $a_{4}$  coefficients for this region of the quasicontinuum are $a_{2} = -0.12 \pm 0.1 (stat) \pm (0.06) (sys)$, and $a_{4} = 0.0 \pm 0.1 (stat) \pm (0.05) (sys)$, where the systematic uncertainties of the quasicontinuum are estimated from the uncertainty in the normalization. 

The uncertainty is large, partly due to the fact that the distribution is a ratio, however the observed value for $a_{2}$ is consistent with a previous measurement at similar energies \cite{PhysRevLett.111.242504}. The absence of $a_{4}$ shows that the data are consistent with pure dipole transitions. It should be noted that in that work, the contributions from stretched quadrupole transitions were estimated to be around 10$\%$ \cite{PhysRevLett.111.242504}. 


Polarization information can be obtained by measuring the angle $\xi$ between the reaction plane, defined by the direction of the photon and the beam-direction, and the Compton scattering plane, defined by the first two Compton scattering interactions in GRETINA. Electric polarization can be discerned from magnetic by constructing an asymmetry parameter defined as:
\[ A = \frac{W(\xi)_{pol}}{W(\xi)_{unpol}}, \]
where $W(\xi)_{pol}$ and $W(\xi)_{unpol}$ are the intensities as a function of the angle $\xi$ for the polarized $\gamma$-rays of interest and a source measurement.  This technique is described in detail by Alikhani \emph{et al.} \cite{ALIKHANI2012144} and Ref. \cite{AWiens}. The effectiveness of GRETINA as a compton polarimeter is demonstrated with two photopeaks in $^{56}$Fe and $^{55}$Fe in Figure \ref{fig:concept}(c) and (d).   The asymmetry $A$ can be expressed in terms of the analyzing power and the degree of linear polarization $P(\theta)$  \cite{ALIKHANI2012144}:

\[ A = \frac{1}{2}Q(E_{\gamma})~P(\theta)cos(2\xi) = \mathcal{A}_{0}~cos(2\xi), \]

\noindent where the $Q(E_{\gamma}$ is the analyzing power. 
The asymmetry $A$ is maximum when $P(\theta)$ is maximum which occurs at $90^{\circ}$ and can be expressed in terms of the angular distribution -- for an E2 transition \cite{RevModPhys.31.711}:

\[ P(\theta)^{E2} = \frac{12 a_{2} + 5 a_{4}}{8 - 4a_{2} + 3a_{4}} \]

Using the $a_{2}$ and $a_{4}$ values for the first-excited state of $^{56}$Fe, the maximum linear polarization is $P = 0.37(11)$, giving an expected asymmetry of $\mathcal{A}_{0} = -0.059(18)$ which agrees well with the observed value of $\mathcal{A}_{E2} = -0.05(1)$, shown in the top two panels of Figure 3. In addition to being consistent with the observed angular distributions,  magnetic polarization (931 keV, $M1$ $^{55}$Fe) is clearly distinguished from electric (847 keV, $E2$ $^{56}$Fe).

\begin{figure}[t!]
\centering
\includegraphics[trim=5 0 0 0, clip, width=1.0\linewidth]{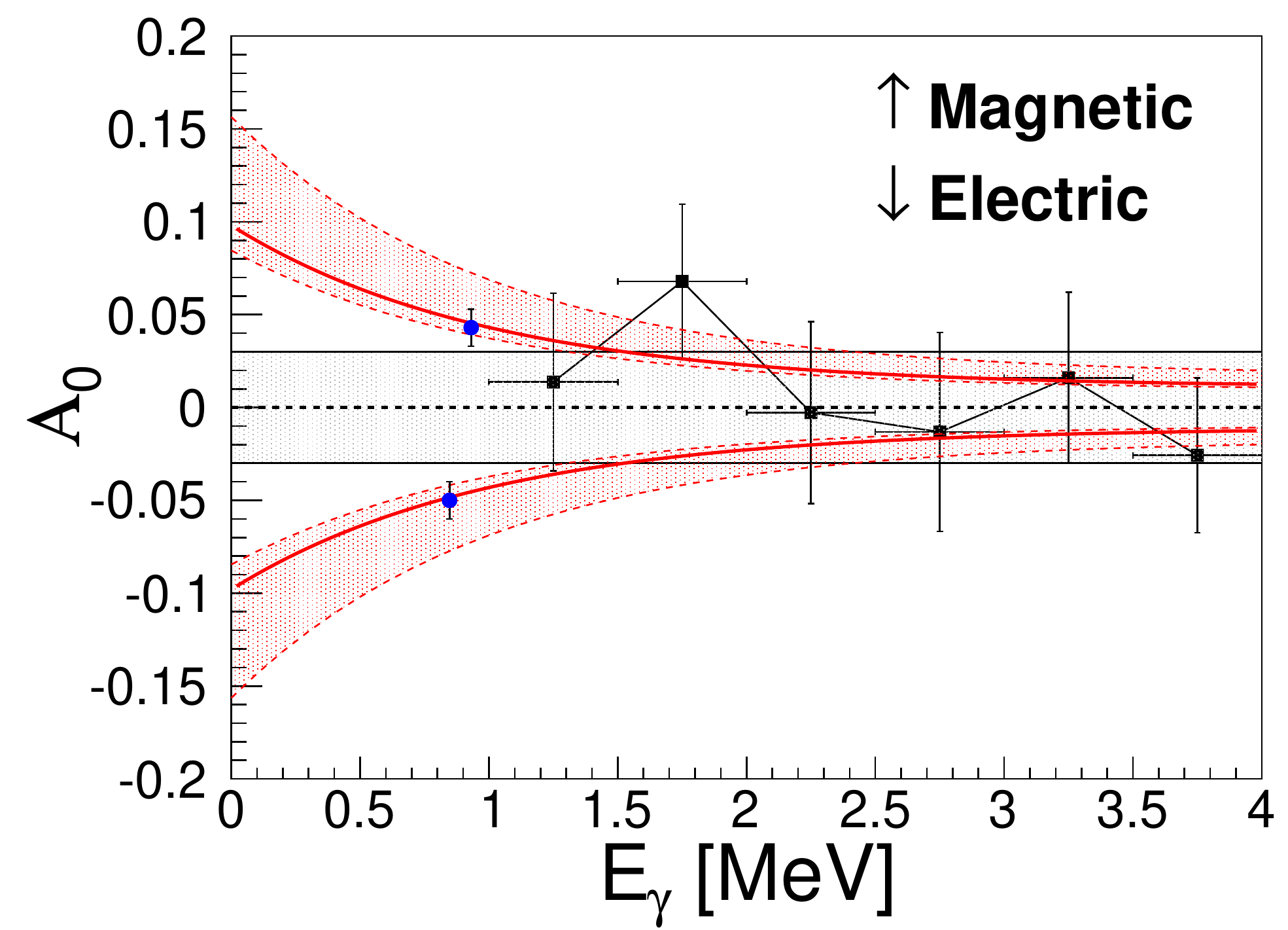}
\caption{(Color online) Polarization asymmetry $\mathcal{A}_{0}$ as a function of primary $\gamma$-ray energy. In blue (circles) are the extracted asymmetries from the best-fits in Figure \ref{fig:concept} for electric and magnetic transitions in $^{55}$Fe and $^{56}$Fe. The red-solid curves denote the expected asymmetry given a linear polarization of $P=0.30$, and the red band (dashed) shows the range of asymmetries given the uncertainty in $P$. The grey band denotes the statistical uncertainty of measuring a uniform distribution. }
\label{fig:asym}
\end{figure}

The polarization asymmetry as a function of primary $\gamma$-ray energy is shown in Figure \ref{fig:asym}, where a negative value indicates electric, and  positive magnetic character. The observed asymmetry for the aforementioned $M1$ and $E2$ transitions are shown in blue with their respective uncertainties for comparison. The red curves show the maximum asymmetry given a linear polarization of $P=0.30$, and the red band (dashed) shows the envelope for the maximum asymmetry given the uncertainty in $P$. The polarization asymmetry for the statistical feeding is shown by the black points, where bin widths of 500 keV are taken.  The error bars represent the statistical limits determined by a $\chi^{2}$ analysis, and the grey band shows the uncertainty in $\mathcal{A}_{0}$ obtained by fitting poisson fluctions of a uniform distribution with identical statistics as the measurement. 


At present, the uncertainties are too large to draw a firm conclusion about a pure electric or magnetic nature of the low-energy enhancement, however the data suggest a mixture, with a small magnetic bias at an observed asymetry of $\mathcal{A}_{0} = 0.06(3)$ in the 1.5 - 2.0 MeV bin. The data are consistent with a uniform distribution within $1\sigma$.  A $\chi^{2}$ hypothesis test shows that this bin is inconsistent with the expected E1 amplitude ($A_{0}$ = -0.03), within $71\%$ confidence, and that $A_{0}$ is not the opposite sign with $87\%$ confidence ($A_{0}$ = -0.06). 

 Given the observed alignment of the $(p,p')$ reaction, the expected asymmetry at $E_{\gamma} > 2$ MeV is $\mathcal{A}_{0} \le 0.03$, which is too small to extract with the present data. In order to enhance the asymmetry it is necessary to probe lower energies where the sensitivity is improved. 

\section{Conclusions}

In summary, the $\gamma$SF of $^{56}$Fe was measured using a $(p,p')$ reaction at 16~MeV, and extracted with the model-independent ratio method. The existence of a low-energy enhancement is confirmed and appears to increase below 1~MeV. In addition, the $\gamma$SFs constructed with $2^{+}$ and $4^{+}$ final states have identical shapes, consistent with the Brink hypothesis. 
The angular distribution is consistent with dipole radiation with an  $a_{2}$ of $-0.12 \pm 0.1 (stat) \pm 0.06 (sys)$. In addition, the polarization asymmetry suggests a mixture of electric and magnetic radiation with a small magnetic bias between 1.5  and 2.0 MeV, however the significance is weak. It will be critical for future experiments to extend polarization measurements to the lowest energies where the sensitivity is the greatest. 

This material is based upon work supported by the U.S. Department of Energy, Office of Science, Office of Nuclear Physics under Contracts No. DE-AC02-05CH11231 (LBNL) , DESC0014442 (WU), DE-AC52-07NA27344 (LLNL), DENA0002905 (OU), and No. DE-AC02-06CH11357 (ANL) and the National Research Foundation of South Africa under Grants No. 92789 and 83867. This research used resources of ANL's ATLAS facility, which is a DOE Office of Science User Facility. A.C.L. gratefully acknowledges funding from ERC-STG-2014 under Grant Agreement No 637686.
\bibliography{library}

\begin{thebibliography}{43}%
\makeatletter
\providecommand \@ifxundefined [1]{%
 \@ifx{#1\undefined}
}%
\providecommand \@ifnum [1]{%
 \ifnum #1\expandafter \@firstoftwo
 \else \expandafter \@secondoftwo
 \fi
}%
\providecommand \@ifx [1]{%
 \ifx #1\expandafter \@firstoftwo
 \else \expandafter \@secondoftwo
 \fi
}%
\providecommand \natexlab [1]{#1}%
\providecommand \enquote  [1]{``#1''}%
\providecommand \bibnamefont  [1]{#1}%
\providecommand \bibfnamefont [1]{#1}%
\providecommand \citenamefont [1]{#1}%
\providecommand \href@noop [0]{\@secondoftwo}%
\providecommand \href [0]{\begingroup \@sanitize@url \@href}%
\providecommand \@href[1]{\@@startlink{#1}\@@href}%
\providecommand \@@href[1]{\endgroup#1\@@endlink}%
\providecommand \@sanitize@url [0]{\catcode `\\12\catcode `\$12\catcode
  `\&12\catcode `\#12\catcode `\^12\catcode `\_12\catcode `\%12\relax}%
\providecommand \@@startlink[1]{}%
\providecommand \@@endlink[0]{}%
\providecommand \url  [0]{\begingroup\@sanitize@url \@url }%
\providecommand \@url [1]{\endgroup\@href {#1}{\urlprefix }}%
\providecommand \urlprefix  [0]{URL }%
\providecommand \Eprint [0]{\href }%
\providecommand \doibase [0]{http://dx.doi.org/}%
\providecommand \selectlanguage [0]{\@gobble}%
\providecommand \bibinfo  [0]{\@secondoftwo}%
\providecommand \bibfield  [0]{\@secondoftwo}%
\providecommand \translation [1]{[#1]}%
\providecommand \BibitemOpen [0]{}%
\providecommand \bibitemStop [0]{}%
\providecommand \bibitemNoStop [0]{.\EOS\space}%
\providecommand \EOS [0]{\spacefactor3000\relax}%
\providecommand \BibitemShut  [1]{\csname bibitem#1\endcsname}%
\let\auto@bib@innerbib\@empty
\bibitem [{\citenamefont {Bartholomew}\ \emph {et~al.}(1973)\citenamefont
  {Bartholomew}, \citenamefont {Earle}, \citenamefont {Ferguson}, \citenamefont
  {Knowles},\ and\ \citenamefont {Lone}}]{Bar73}%
  \BibitemOpen
  \bibfield  {author} {\bibinfo {author} {\bibfnamefont {G.~A.}\ \bibnamefont
  {Bartholomew}}, \bibinfo {author} {\bibfnamefont {E.~D.}\ \bibnamefont
  {Earle}}, \bibinfo {author} {\bibfnamefont {A.~J.}\ \bibnamefont {Ferguson}},
  \bibinfo {author} {\bibfnamefont {J.~W.}\ \bibnamefont {Knowles}}, \ and\
  \bibinfo {author} {\bibfnamefont {M.~A.}\ \bibnamefont {Lone}},\ }in\ \href
  {\doibase 10.1007/978-1-4615-9044-6_4} {\emph {\bibinfo {booktitle} {Adv.
  Nucl. Phys. 7, 229}}}\ (\bibinfo {year} {1973})\BibitemShut {NoStop}%
\bibitem [{\citenamefont {Harakeh}\ and\ \citenamefont {van~der
  Woude}(2001)}]{Har01}%
  \BibitemOpen
  \bibfield  {author} {\bibinfo {author} {\bibfnamefont {M.~N.}\ \bibnamefont
  {Harakeh}}\ and\ \bibinfo {author} {\bibfnamefont {A.}~\bibnamefont {van~der
  Woude}},\ }\href@noop {} {\bibfield  {journal} {\bibinfo  {journal} {Oxford
  University Press, Oxford}\ } (\bibinfo {year} {2001})}\BibitemShut {NoStop}%
\bibitem [{\citenamefont {Bracco}\ \emph {et~al.}(2015)\citenamefont {Bracco},
  \citenamefont {Crespi},\ and\ \citenamefont {Lanza}}]{Bra15}%
  \BibitemOpen
  \bibfield  {author} {\bibinfo {author} {\bibfnamefont {A.}~\bibnamefont
  {Bracco}}, \bibinfo {author} {\bibfnamefont {F.~C.~L.}\ \bibnamefont
  {Crespi}}, \ and\ \bibinfo {author} {\bibfnamefont {E.~G.}\ \bibnamefont
  {Lanza}},\ }\href {\doibase 10.1140/epja/i2015-15099-6} {\bibfield  {journal}
  {\bibinfo  {journal} {The European Physical Journal A}\ }\textbf {\bibinfo
  {volume} {51}},\ \bibinfo {pages} {99} (\bibinfo {year} {2015})}\BibitemShut
  {NoStop}%
\bibitem [{\citenamefont {Savran}\ \emph {et~al.}(2013)\citenamefont {Savran},
  \citenamefont {Aumann},\ and\ \citenamefont {Zilges}}]{SAVRAN2013210}%
  \BibitemOpen
  \bibfield  {author} {\bibinfo {author} {\bibfnamefont {D.}~\bibnamefont
  {Savran}}, \bibinfo {author} {\bibfnamefont {T.}~\bibnamefont {Aumann}}, \
  and\ \bibinfo {author} {\bibfnamefont {A.}~\bibnamefont {Zilges}},\ }\href
  {\doibase https://doi.org/10.1016/j.ppnp.2013.02.003} {\bibfield  {journal}
  {\bibinfo  {journal} {Progress in Particle and Nuclear Physics}\ }\textbf
  {\bibinfo {volume} {70}},\ \bibinfo {pages} {210 } (\bibinfo {year}
  {2013})}\BibitemShut {NoStop}%
\bibitem [{\citenamefont {Schiller}\ \emph {et~al.}(2006)\citenamefont
  {Schiller}, \citenamefont {Voinov}, \citenamefont {Algin}, \citenamefont
  {Becker}, \citenamefont {Bernstein}, \citenamefont {Garrett}, \citenamefont
  {Guttormsen}, \citenamefont {Nelson}, \citenamefont {Rekstad},\ and\
  \citenamefont {Siem}}]{Sch06}%
  \BibitemOpen
  \bibfield  {author} {\bibinfo {author} {\bibfnamefont {A.}~\bibnamefont
  {Schiller}}, \bibinfo {author} {\bibfnamefont {A.}~\bibnamefont {Voinov}},
  \bibinfo {author} {\bibfnamefont {E.}~\bibnamefont {Algin}}, \bibinfo
  {author} {\bibfnamefont {J.}~\bibnamefont {Becker}}, \bibinfo {author}
  {\bibfnamefont {L.}~\bibnamefont {Bernstein}}, \bibinfo {author}
  {\bibfnamefont {P.}~\bibnamefont {Garrett}}, \bibinfo {author} {\bibfnamefont
  {M.}~\bibnamefont {Guttormsen}}, \bibinfo {author} {\bibfnamefont
  {R.}~\bibnamefont {Nelson}}, \bibinfo {author} {\bibfnamefont
  {J.}~\bibnamefont {Rekstad}}, \ and\ \bibinfo {author} {\bibfnamefont
  {S.}~\bibnamefont {Siem}},\ }\href {\doibase
  http://dx.doi.org/10.1016/j.physletb.2005.12.043} {\bibfield  {journal}
  {\bibinfo  {journal} {Physics Letters B}\ }\textbf {\bibinfo {volume}
  {633}},\ \bibinfo {pages} {225 } (\bibinfo {year} {2006})}\BibitemShut
  {NoStop}%
\bibitem [{\citenamefont {Heyde}\ \emph {et~al.}(2010)\citenamefont {Heyde},
  \citenamefont {von Neumann-Cosel},\ and\ \citenamefont {Richter}}]{Neu10}%
  \BibitemOpen
  \bibfield  {author} {\bibinfo {author} {\bibfnamefont {K.}~\bibnamefont
  {Heyde}}, \bibinfo {author} {\bibfnamefont {P.}~\bibnamefont {von
  Neumann-Cosel}}, \ and\ \bibinfo {author} {\bibfnamefont {A.}~\bibnamefont
  {Richter}},\ }\href {\doibase 10.1103/RevModPhys.82.2365} {\bibfield
  {journal} {\bibinfo  {journal} {Rev. Mod. Phys.}\ }\textbf {\bibinfo {volume}
  {82}},\ \bibinfo {pages} {2365} (\bibinfo {year} {2010})}\BibitemShut
  {NoStop}%
\bibitem [{\citenamefont {Goriely}(1998)}]{GORIELY199810}%
  \BibitemOpen
  \bibfield  {author} {\bibinfo {author} {\bibfnamefont {S.}~\bibnamefont
  {Goriely}},\ }\href {\doibase
  http://dx.doi.org/10.1016/S0370-2693(98)00907-1} {\bibfield  {journal}
  {\bibinfo  {journal} {Physics Letters B}\ }\textbf {\bibinfo {volume}
  {436}},\ \bibinfo {pages} {10 } (\bibinfo {year} {1998})}\BibitemShut
  {NoStop}%
\bibitem [{\citenamefont {Hauser}\ and\ \citenamefont
  {Feshbach}(1952)}]{Hau52}%
  \BibitemOpen
  \bibfield  {author} {\bibinfo {author} {\bibfnamefont {W.}~\bibnamefont
  {Hauser}}\ and\ \bibinfo {author} {\bibfnamefont {H.}~\bibnamefont
  {Feshbach}},\ }\href {\doibase 10.1103/PhysRev.87.366} {\bibfield  {journal}
  {\bibinfo  {journal} {Phys. Rev.}\ }\textbf {\bibinfo {volume} {87}},\
  \bibinfo {pages} {366} (\bibinfo {year} {1952})}\BibitemShut {NoStop}%
\bibitem [{\citenamefont {Arnould}\ and\ \citenamefont
  {Goriely}(2003)}]{Arn03}%
  \BibitemOpen
  \bibfield  {author} {\bibinfo {author} {\bibfnamefont {M.}~\bibnamefont
  {Arnould}}\ and\ \bibinfo {author} {\bibfnamefont {S.}~\bibnamefont
  {Goriely}},\ }\href {\doibase
  http://dx.doi.org/10.1016/S0370-1573(03)00242-4} {\bibfield  {journal}
  {\bibinfo  {journal} {Physics Reports}\ }\textbf {\bibinfo {volume} {384}},\
  \bibinfo {pages} {1 } (\bibinfo {year} {2003})}\BibitemShut {NoStop}%
\bibitem [{\citenamefont {Arnould}\ \emph {et~al.}(2007)\citenamefont
  {Arnould}, \citenamefont {Goriely},\ and\ \citenamefont {Takahashi}}]{Arn07}%
  \BibitemOpen
  \bibfield  {author} {\bibinfo {author} {\bibfnamefont {M.}~\bibnamefont
  {Arnould}}, \bibinfo {author} {\bibfnamefont {S.}~\bibnamefont {Goriely}}, \
  and\ \bibinfo {author} {\bibfnamefont {K.}~\bibnamefont {Takahashi}},\ }\href
  {\doibase http://dx.doi.org/10.1016/j.physrep.2007.06.002} {\bibfield
  {journal} {\bibinfo  {journal} {Physics Reports}\ }\textbf {\bibinfo {volume}
  {450}},\ \bibinfo {pages} {97 } (\bibinfo {year} {2007})}\BibitemShut
  {NoStop}%
\bibitem [{\citenamefont {Colonna}\ \emph {et~al.}(2010)\citenamefont
  {Colonna}, \citenamefont {Belloni}, \citenamefont {Berthoumieux},
  \citenamefont {Calviani}, \citenamefont {Domingo-Pardo}, \citenamefont
  {Guerrero}, \citenamefont {Karadimos}, \citenamefont {Lederer}, \citenamefont
  {Massimi}, \citenamefont {Paradela}, \citenamefont {Plag}, \citenamefont
  {Praena},\ and\ \citenamefont {Sarmento}}]{Col10}%
  \BibitemOpen
  \bibfield  {author} {\bibinfo {author} {\bibfnamefont {N.}~\bibnamefont
  {Colonna}}, \bibinfo {author} {\bibfnamefont {F.}~\bibnamefont {Belloni}},
  \bibinfo {author} {\bibfnamefont {E.}~\bibnamefont {Berthoumieux}}, \bibinfo
  {author} {\bibfnamefont {M.}~\bibnamefont {Calviani}}, \bibinfo {author}
  {\bibfnamefont {C.}~\bibnamefont {Domingo-Pardo}}, \bibinfo {author}
  {\bibfnamefont {C.}~\bibnamefont {Guerrero}}, \bibinfo {author}
  {\bibfnamefont {D.}~\bibnamefont {Karadimos}}, \bibinfo {author}
  {\bibfnamefont {C.}~\bibnamefont {Lederer}}, \bibinfo {author} {\bibfnamefont
  {C.}~\bibnamefont {Massimi}}, \bibinfo {author} {\bibfnamefont
  {C.}~\bibnamefont {Paradela}}, \bibinfo {author} {\bibfnamefont
  {R.}~\bibnamefont {Plag}}, \bibinfo {author} {\bibfnamefont {J.}~\bibnamefont
  {Praena}}, \ and\ \bibinfo {author} {\bibfnamefont {R.}~\bibnamefont
  {Sarmento}},\ }\href {\doibase 10.1039/C0EE00108B} {\bibfield  {journal}
  {\bibinfo  {journal} {Energy Environ. Sci.}\ }\textbf {\bibinfo {volume}
  {3}},\ \bibinfo {pages} {1910} (\bibinfo {year} {2010})}\BibitemShut
  {NoStop}%
\bibitem [{Rep(2006)}]{Report06}%
  \BibitemOpen
  \href@noop {} {\bibfield  {journal} {\bibinfo  {journal} {Report of the
  Nuclear Physics and Related Computational Science R\&D for Advanced Fuel
  Cycles Workshop, DOE Offices of Nuclear Physics and Advanced Scientific
  Computing Research.}\ } (\bibinfo {year} {August 2006})}\BibitemShut
  {NoStop}%
\bibitem [{\citenamefont {Larsen}\ and\ \citenamefont
  {Goriely}(2010)}]{PhysRevC.82.014318}%
  \BibitemOpen
  \bibfield  {author} {\bibinfo {author} {\bibfnamefont {A.~C.}\ \bibnamefont
  {Larsen}}\ and\ \bibinfo {author} {\bibfnamefont {S.}~\bibnamefont
  {Goriely}},\ }\href {\doibase 10.1103/PhysRevC.82.014318} {\bibfield
  {journal} {\bibinfo  {journal} {Phys. Rev. C}\ }\textbf {\bibinfo {volume}
  {82}},\ \bibinfo {pages} {014318} (\bibinfo {year} {2010})}\BibitemShut
  {NoStop}%
\bibitem [{\citenamefont {Burbidge}\ \emph {et~al.}(1957)\citenamefont
  {Burbidge}, \citenamefont {Burbidge}, \citenamefont {Fowler},\ and\
  \citenamefont {Hoyle}}]{RevModPhys.29.547}%
  \BibitemOpen
  \bibfield  {author} {\bibinfo {author} {\bibfnamefont {E.~M.}\ \bibnamefont
  {Burbidge}}, \bibinfo {author} {\bibfnamefont {G.~R.}\ \bibnamefont
  {Burbidge}}, \bibinfo {author} {\bibfnamefont {W.~A.}\ \bibnamefont
  {Fowler}}, \ and\ \bibinfo {author} {\bibfnamefont {F.}~\bibnamefont
  {Hoyle}},\ }\href {\doibase 10.1103/RevModPhys.29.547} {\bibfield  {journal}
  {\bibinfo  {journal} {Rev. Mod. Phys.}\ }\textbf {\bibinfo {volume} {29}},\
  \bibinfo {pages} {547} (\bibinfo {year} {1957})}\BibitemShut {NoStop}%
\bibitem [{\citenamefont {Laplace}\ \emph {et~al.}(2016)\citenamefont
  {Laplace}, \citenamefont {Zeiser}, \citenamefont {Guttormsen}, \citenamefont
  {Larsen}, \citenamefont {Bleuel}, \citenamefont {Bernstein}, \citenamefont
  {Goldblum}, \citenamefont {Siem}, \citenamefont {Garotte}, \citenamefont
  {Brown}, \citenamefont {Campo}, \citenamefont {Eriksen}, \citenamefont
  {Giacoppo}, \citenamefont {G\"orgen}, \citenamefont {Hady\ifmmode
  \acute{n}\else \'{n}\fi{}ska-Kl\ifmmode~\mbox{\c{e}}\else \c{e}\fi{}k},
  \citenamefont {Henderson}, \citenamefont {Klintefjord}, \citenamefont
  {Lebois}, \citenamefont {Renstr\o{}m}, \citenamefont {Rose}, \citenamefont
  {Sahin}, \citenamefont {Tornyi}, \citenamefont {Tveten}, \citenamefont
  {Voinov}, \citenamefont {Wiedeking}, \citenamefont {Wilson},\ and\
  \citenamefont {Younes}}]{Lap16}%
  \BibitemOpen
  \bibfield  {author} {\bibinfo {author} {\bibfnamefont {T.~A.}\ \bibnamefont
  {Laplace}}, \bibinfo {author} {\bibfnamefont {F.}~\bibnamefont {Zeiser}},
  \bibinfo {author} {\bibfnamefont {M.}~\bibnamefont {Guttormsen}}, \bibinfo
  {author} {\bibfnamefont {A.~C.}\ \bibnamefont {Larsen}}, \bibinfo {author}
  {\bibfnamefont {D.~L.}\ \bibnamefont {Bleuel}}, \bibinfo {author}
  {\bibfnamefont {L.~A.}\ \bibnamefont {Bernstein}}, \bibinfo {author}
  {\bibfnamefont {B.~L.}\ \bibnamefont {Goldblum}}, \bibinfo {author}
  {\bibfnamefont {S.}~\bibnamefont {Siem}}, \bibinfo {author} {\bibfnamefont
  {F.~L.~B.}\ \bibnamefont {Garotte}}, \bibinfo {author} {\bibfnamefont
  {J.~A.}\ \bibnamefont {Brown}}, \bibinfo {author} {\bibfnamefont {L.~C.}\
  \bibnamefont {Campo}}, \bibinfo {author} {\bibfnamefont {T.~K.}\ \bibnamefont
  {Eriksen}}, \bibinfo {author} {\bibfnamefont {F.}~\bibnamefont {Giacoppo}},
  \bibinfo {author} {\bibfnamefont {A.}~\bibnamefont {G\"orgen}}, \bibinfo
  {author} {\bibfnamefont {K.}~\bibnamefont {Hady\ifmmode \acute{n}\else
  \'{n}\fi{}ska-Kl\ifmmode~\mbox{\c{e}}\else \c{e}\fi{}k}}, \bibinfo {author}
  {\bibfnamefont {R.~A.}\ \bibnamefont {Henderson}}, \bibinfo {author}
  {\bibfnamefont {M.}~\bibnamefont {Klintefjord}}, \bibinfo {author}
  {\bibfnamefont {M.}~\bibnamefont {Lebois}}, \bibinfo {author} {\bibfnamefont
  {T.}~\bibnamefont {Renstr\o{}m}}, \bibinfo {author} {\bibfnamefont {S.~J.}\
  \bibnamefont {Rose}}, \bibinfo {author} {\bibfnamefont {E.}~\bibnamefont
  {Sahin}}, \bibinfo {author} {\bibfnamefont {T.~G.}\ \bibnamefont {Tornyi}},
  \bibinfo {author} {\bibfnamefont {G.~M.}\ \bibnamefont {Tveten}}, \bibinfo
  {author} {\bibfnamefont {A.}~\bibnamefont {Voinov}}, \bibinfo {author}
  {\bibfnamefont {M.}~\bibnamefont {Wiedeking}}, \bibinfo {author}
  {\bibfnamefont {J.~N.}\ \bibnamefont {Wilson}}, \ and\ \bibinfo {author}
  {\bibfnamefont {W.}~\bibnamefont {Younes}},\ }\href {\doibase
  10.1103/PhysRevC.93.014323} {\bibfield  {journal} {\bibinfo  {journal} {Phys.
  Rev. C}\ }\textbf {\bibinfo {volume} {93}},\ \bibinfo {pages} {014323}
  (\bibinfo {year} {2016})}\BibitemShut {NoStop}%
\bibitem [{\citenamefont {Kheswa}\ \emph {et~al.}(2017)\citenamefont {Kheswa},
  \citenamefont {Wiedeking}, \citenamefont {Brown}, \citenamefont {Larsen},
  \citenamefont {Goriely}, \citenamefont {Guttormsen}, \citenamefont
  {Bello~Garrote}, \citenamefont {Bernstein}, \citenamefont {Bleuel},
  \citenamefont {Eriksen}, \citenamefont {Giacoppo}, \citenamefont {G\"orgen},
  \citenamefont {Goldblum}, \citenamefont {Hagen}, \citenamefont {Koehler},
  \citenamefont {Klintefjord}, \citenamefont {Malatji}, \citenamefont
  {Midtb\o{}}, \citenamefont {Nyhus}, \citenamefont {Papka}, \citenamefont
  {Renstr\o{}m}, \citenamefont {Rose}, \citenamefont {Sahin}, \citenamefont
  {Siem},\ and\ \citenamefont {Tornyi}}]{Khe17}%
  \BibitemOpen
  \bibfield  {author} {\bibinfo {author} {\bibfnamefont {B.~V.}\ \bibnamefont
  {Kheswa}}, \bibinfo {author} {\bibfnamefont {M.}~\bibnamefont {Wiedeking}},
  \bibinfo {author} {\bibfnamefont {J.~A.}\ \bibnamefont {Brown}}, \bibinfo
  {author} {\bibfnamefont {A.~C.}\ \bibnamefont {Larsen}}, \bibinfo {author}
  {\bibfnamefont {S.}~\bibnamefont {Goriely}}, \bibinfo {author} {\bibfnamefont
  {M.}~\bibnamefont {Guttormsen}}, \bibinfo {author} {\bibfnamefont {F.~L.}\
  \bibnamefont {Bello~Garrote}}, \bibinfo {author} {\bibfnamefont {L.~A.}\
  \bibnamefont {Bernstein}}, \bibinfo {author} {\bibfnamefont {D.~L.}\
  \bibnamefont {Bleuel}}, \bibinfo {author} {\bibfnamefont {T.~K.}\
  \bibnamefont {Eriksen}}, \bibinfo {author} {\bibfnamefont {F.}~\bibnamefont
  {Giacoppo}}, \bibinfo {author} {\bibfnamefont {A.}~\bibnamefont {G\"orgen}},
  \bibinfo {author} {\bibfnamefont {B.~L.}\ \bibnamefont {Goldblum}}, \bibinfo
  {author} {\bibfnamefont {T.~W.}\ \bibnamefont {Hagen}}, \bibinfo {author}
  {\bibfnamefont {P.~E.}\ \bibnamefont {Koehler}}, \bibinfo {author}
  {\bibfnamefont {M.}~\bibnamefont {Klintefjord}}, \bibinfo {author}
  {\bibfnamefont {K.~L.}\ \bibnamefont {Malatji}}, \bibinfo {author}
  {\bibfnamefont {J.~E.}\ \bibnamefont {Midtb\o{}}}, \bibinfo {author}
  {\bibfnamefont {H.~T.}\ \bibnamefont {Nyhus}}, \bibinfo {author}
  {\bibfnamefont {P.}~\bibnamefont {Papka}}, \bibinfo {author} {\bibfnamefont
  {T.}~\bibnamefont {Renstr\o{}m}}, \bibinfo {author} {\bibfnamefont {S.~J.}\
  \bibnamefont {Rose}}, \bibinfo {author} {\bibfnamefont {E.}~\bibnamefont
  {Sahin}}, \bibinfo {author} {\bibfnamefont {S.}~\bibnamefont {Siem}}, \ and\
  \bibinfo {author} {\bibfnamefont {T.~G.}\ \bibnamefont {Tornyi}},\ }\href
  {\doibase 10.1103/PhysRevC.95.045805} {\bibfield  {journal} {\bibinfo
  {journal} {Phys. Rev. C}\ }\textbf {\bibinfo {volume} {95}},\ \bibinfo
  {pages} {045805} (\bibinfo {year} {2017})}\BibitemShut {NoStop}%
\bibitem [{\citenamefont {Larsen}\ \emph {et~al.}(2016)\citenamefont {Larsen},
  \citenamefont {Guttormsen}, \citenamefont {Schwengner}, \citenamefont
  {Bleuel}, \citenamefont {Goriely}, \citenamefont {Harissopulos},
  \citenamefont {Bello~Garrote}, \citenamefont {Byun}, \citenamefont {Eriksen},
  \citenamefont {Giacoppo}, \citenamefont {G\"orgen}, \citenamefont {Hagen},
  \citenamefont {Klintefjord}, \citenamefont {Renstr\o{}m}, \citenamefont
  {Rose}, \citenamefont {Sahin}, \citenamefont {Siem}, \citenamefont {Tornyi},
  \citenamefont {Tveten}, \citenamefont {Voinov},\ and\ \citenamefont
  {Wiedeking}}]{Lar16}%
  \BibitemOpen
  \bibfield  {author} {\bibinfo {author} {\bibfnamefont {A.~C.}\ \bibnamefont
  {Larsen}}, \bibinfo {author} {\bibfnamefont {M.}~\bibnamefont {Guttormsen}},
  \bibinfo {author} {\bibfnamefont {R.}~\bibnamefont {Schwengner}}, \bibinfo
  {author} {\bibfnamefont {D.~L.}\ \bibnamefont {Bleuel}}, \bibinfo {author}
  {\bibfnamefont {S.}~\bibnamefont {Goriely}}, \bibinfo {author} {\bibfnamefont
  {S.}~\bibnamefont {Harissopulos}}, \bibinfo {author} {\bibfnamefont {F.~L.}\
  \bibnamefont {Bello~Garrote}}, \bibinfo {author} {\bibfnamefont
  {Y.}~\bibnamefont {Byun}}, \bibinfo {author} {\bibfnamefont {T.~K.}\
  \bibnamefont {Eriksen}}, \bibinfo {author} {\bibfnamefont {F.}~\bibnamefont
  {Giacoppo}}, \bibinfo {author} {\bibfnamefont {A.}~\bibnamefont {G\"orgen}},
  \bibinfo {author} {\bibfnamefont {T.~W.}\ \bibnamefont {Hagen}}, \bibinfo
  {author} {\bibfnamefont {M.}~\bibnamefont {Klintefjord}}, \bibinfo {author}
  {\bibfnamefont {T.}~\bibnamefont {Renstr\o{}m}}, \bibinfo {author}
  {\bibfnamefont {S.~J.}\ \bibnamefont {Rose}}, \bibinfo {author}
  {\bibfnamefont {E.}~\bibnamefont {Sahin}}, \bibinfo {author} {\bibfnamefont
  {S.}~\bibnamefont {Siem}}, \bibinfo {author} {\bibfnamefont {T.~G.}\
  \bibnamefont {Tornyi}}, \bibinfo {author} {\bibfnamefont {G.~M.}\
  \bibnamefont {Tveten}}, \bibinfo {author} {\bibfnamefont {A.~V.}\
  \bibnamefont {Voinov}}, \ and\ \bibinfo {author} {\bibfnamefont
  {M.}~\bibnamefont {Wiedeking}},\ }\href {\doibase 10.1103/PhysRevC.93.045810}
  {\bibfield  {journal} {\bibinfo  {journal} {Phys. Rev. C}\ }\textbf {\bibinfo
  {volume} {93}},\ \bibinfo {pages} {045810} (\bibinfo {year}
  {2016})}\BibitemShut {NoStop}%
\bibitem [{\citenamefont {Spyrou}\ \emph {et~al.}(2014)\citenamefont {Spyrou},
  \citenamefont {Liddick}, \citenamefont {Larsen}, \citenamefont {Guttormsen},
  \citenamefont {Cooper}, \citenamefont {Dombos}, \citenamefont {Morrissey},
  \citenamefont {Naqvi}, \citenamefont {Perdikakis}, \citenamefont {Quinn},
  \citenamefont {Renstr\o{}m}, \citenamefont {Rodriguez}, \citenamefont
  {Simon}, \citenamefont {Sumithrarachchi},\ and\ \citenamefont
  {Zegers}}]{Spy14}%
  \BibitemOpen
  \bibfield  {author} {\bibinfo {author} {\bibfnamefont {A.}~\bibnamefont
  {Spyrou}}, \bibinfo {author} {\bibfnamefont {S.~N.}\ \bibnamefont {Liddick}},
  \bibinfo {author} {\bibfnamefont {A.~C.}\ \bibnamefont {Larsen}}, \bibinfo
  {author} {\bibfnamefont {M.}~\bibnamefont {Guttormsen}}, \bibinfo {author}
  {\bibfnamefont {K.}~\bibnamefont {Cooper}}, \bibinfo {author} {\bibfnamefont
  {A.~C.}\ \bibnamefont {Dombos}}, \bibinfo {author} {\bibfnamefont {D.~J.}\
  \bibnamefont {Morrissey}}, \bibinfo {author} {\bibfnamefont {F.}~\bibnamefont
  {Naqvi}}, \bibinfo {author} {\bibfnamefont {G.}~\bibnamefont {Perdikakis}},
  \bibinfo {author} {\bibfnamefont {S.~J.}\ \bibnamefont {Quinn}}, \bibinfo
  {author} {\bibfnamefont {T.}~\bibnamefont {Renstr\o{}m}}, \bibinfo {author}
  {\bibfnamefont {J.~A.}\ \bibnamefont {Rodriguez}}, \bibinfo {author}
  {\bibfnamefont {A.}~\bibnamefont {Simon}}, \bibinfo {author} {\bibfnamefont
  {C.~S.}\ \bibnamefont {Sumithrarachchi}}, \ and\ \bibinfo {author}
  {\bibfnamefont {R.~G.~T.}\ \bibnamefont {Zegers}},\ }\href {\doibase
  10.1103/PhysRevLett.113.232502} {\bibfield  {journal} {\bibinfo  {journal}
  {Phys. Rev. Lett.}\ }\textbf {\bibinfo {volume} {113}},\ \bibinfo {pages}
  {232502} (\bibinfo {year} {2014})}\BibitemShut {NoStop}%
\bibitem [{\citenamefont {Liddick}\ \emph {et~al.}(2016)\citenamefont
  {Liddick}, \citenamefont {Spyrou}, \citenamefont {Crider}, \citenamefont
  {Naqvi}, \citenamefont {Larsen}, \citenamefont {Guttormsen}, \citenamefont
  {Mumpower}, \citenamefont {Surman}, \citenamefont {Perdikakis}, \citenamefont
  {Bleuel}, \citenamefont {Couture}, \citenamefont {Crespo~Campo},
  \citenamefont {Dombos}, \citenamefont {Lewis}, \citenamefont {Mosby},
  \citenamefont {Nikas}, \citenamefont {Prokop}, \citenamefont {Renstrom},
  \citenamefont {Rubio}, \citenamefont {Siem},\ and\ \citenamefont
  {Quinn}}]{Lid16}%
  \BibitemOpen
  \bibfield  {author} {\bibinfo {author} {\bibfnamefont {S.~N.}\ \bibnamefont
  {Liddick}}, \bibinfo {author} {\bibfnamefont {A.}~\bibnamefont {Spyrou}},
  \bibinfo {author} {\bibfnamefont {B.~P.}\ \bibnamefont {Crider}}, \bibinfo
  {author} {\bibfnamefont {F.}~\bibnamefont {Naqvi}}, \bibinfo {author}
  {\bibfnamefont {A.~C.}\ \bibnamefont {Larsen}}, \bibinfo {author}
  {\bibfnamefont {M.}~\bibnamefont {Guttormsen}}, \bibinfo {author}
  {\bibfnamefont {M.}~\bibnamefont {Mumpower}}, \bibinfo {author}
  {\bibfnamefont {R.}~\bibnamefont {Surman}}, \bibinfo {author} {\bibfnamefont
  {G.}~\bibnamefont {Perdikakis}}, \bibinfo {author} {\bibfnamefont {D.~L.}\
  \bibnamefont {Bleuel}}, \bibinfo {author} {\bibfnamefont {A.}~\bibnamefont
  {Couture}}, \bibinfo {author} {\bibfnamefont {L.}~\bibnamefont
  {Crespo~Campo}}, \bibinfo {author} {\bibfnamefont {A.~C.}\ \bibnamefont
  {Dombos}}, \bibinfo {author} {\bibfnamefont {R.}~\bibnamefont {Lewis}},
  \bibinfo {author} {\bibfnamefont {S.}~\bibnamefont {Mosby}}, \bibinfo
  {author} {\bibfnamefont {S.}~\bibnamefont {Nikas}}, \bibinfo {author}
  {\bibfnamefont {C.~J.}\ \bibnamefont {Prokop}}, \bibinfo {author}
  {\bibfnamefont {T.}~\bibnamefont {Renstrom}}, \bibinfo {author}
  {\bibfnamefont {B.}~\bibnamefont {Rubio}}, \bibinfo {author} {\bibfnamefont
  {S.}~\bibnamefont {Siem}}, \ and\ \bibinfo {author} {\bibfnamefont {S.~J.}\
  \bibnamefont {Quinn}},\ }\href {\doibase 10.1103/PhysRevLett.116.242502}
  {\bibfield  {journal} {\bibinfo  {journal} {Phys. Rev. Lett.}\ }\textbf
  {\bibinfo {volume} {116}},\ \bibinfo {pages} {242502} (\bibinfo {year}
  {2016})}\BibitemShut {NoStop}%
\bibitem [{\citenamefont {Kozub}\ \emph {et~al.}(2012)\citenamefont {Kozub},
  \citenamefont {Arbanas}, \citenamefont {Adekola}, \citenamefont {Bardayan},
  \citenamefont {Blackmon}, \citenamefont {Chae}, \citenamefont {Chipps},
  \citenamefont {Cizewski}, \citenamefont {Erikson}, \citenamefont {Hatarik},
  \citenamefont {Hix}, \citenamefont {Jones}, \citenamefont {Krolas},
  \citenamefont {Liang}, \citenamefont {Ma}, \citenamefont {Matei},
  \citenamefont {Moazen}, \citenamefont {Nesaraja}, \citenamefont {Pain},
  \citenamefont {Shapira}, \citenamefont {Shriner}, \citenamefont {Smith},\
  and\ \citenamefont {Swan}}]{PhysRevLett.109.172501}%
  \BibitemOpen
  \bibfield  {author} {\bibinfo {author} {\bibfnamefont {R.~L.}\ \bibnamefont
  {Kozub}}, \bibinfo {author} {\bibfnamefont {G.}~\bibnamefont {Arbanas}},
  \bibinfo {author} {\bibfnamefont {A.~S.}\ \bibnamefont {Adekola}}, \bibinfo
  {author} {\bibfnamefont {D.~W.}\ \bibnamefont {Bardayan}}, \bibinfo {author}
  {\bibfnamefont {J.~C.}\ \bibnamefont {Blackmon}}, \bibinfo {author}
  {\bibfnamefont {K.~Y.}\ \bibnamefont {Chae}}, \bibinfo {author}
  {\bibfnamefont {K.~A.}\ \bibnamefont {Chipps}}, \bibinfo {author}
  {\bibfnamefont {J.~A.}\ \bibnamefont {Cizewski}}, \bibinfo {author}
  {\bibfnamefont {L.}~\bibnamefont {Erikson}}, \bibinfo {author} {\bibfnamefont
  {R.}~\bibnamefont {Hatarik}}, \bibinfo {author} {\bibfnamefont {W.~R.}\
  \bibnamefont {Hix}}, \bibinfo {author} {\bibfnamefont {K.~L.}\ \bibnamefont
  {Jones}}, \bibinfo {author} {\bibfnamefont {W.}~\bibnamefont {Krolas}},
  \bibinfo {author} {\bibfnamefont {J.~F.}\ \bibnamefont {Liang}}, \bibinfo
  {author} {\bibfnamefont {Z.}~\bibnamefont {Ma}}, \bibinfo {author}
  {\bibfnamefont {C.}~\bibnamefont {Matei}}, \bibinfo {author} {\bibfnamefont
  {B.~H.}\ \bibnamefont {Moazen}}, \bibinfo {author} {\bibfnamefont {C.~D.}\
  \bibnamefont {Nesaraja}}, \bibinfo {author} {\bibfnamefont {S.~D.}\
  \bibnamefont {Pain}}, \bibinfo {author} {\bibfnamefont {D.}~\bibnamefont
  {Shapira}}, \bibinfo {author} {\bibfnamefont {J.~F.}\ \bibnamefont
  {Shriner}}, \bibinfo {author} {\bibfnamefont {M.~S.}\ \bibnamefont {Smith}},
  \ and\ \bibinfo {author} {\bibfnamefont {T.~P.}\ \bibnamefont {Swan}},\
  }\href {\doibase 10.1103/PhysRevLett.109.172501} {\bibfield  {journal}
  {\bibinfo  {journal} {Phys. Rev. Lett.}\ }\textbf {\bibinfo {volume} {109}},\
  \bibinfo {pages} {172501} (\bibinfo {year} {2012})}\BibitemShut {NoStop}%
\bibitem [{\citenamefont {Voinov}\ \emph {et~al.}(2004)\citenamefont {Voinov},
  \citenamefont {Algin}, \citenamefont {Agvaanluvsan}, \citenamefont {Belgya},
  \citenamefont {Chankova}, \citenamefont {Guttormsen}, \citenamefont
  {Mitchell}, \citenamefont {Rekstad}, \citenamefont {Schiller},\ and\
  \citenamefont {Siem}}]{PhysRevLett.93.142504}%
  \BibitemOpen
  \bibfield  {author} {\bibinfo {author} {\bibfnamefont {A.}~\bibnamefont
  {Voinov}}, \bibinfo {author} {\bibfnamefont {E.}~\bibnamefont {Algin}},
  \bibinfo {author} {\bibfnamefont {U.}~\bibnamefont {Agvaanluvsan}}, \bibinfo
  {author} {\bibfnamefont {T.}~\bibnamefont {Belgya}}, \bibinfo {author}
  {\bibfnamefont {R.}~\bibnamefont {Chankova}}, \bibinfo {author}
  {\bibfnamefont {M.}~\bibnamefont {Guttormsen}}, \bibinfo {author}
  {\bibfnamefont {G.~E.}\ \bibnamefont {Mitchell}}, \bibinfo {author}
  {\bibfnamefont {J.}~\bibnamefont {Rekstad}}, \bibinfo {author} {\bibfnamefont
  {A.}~\bibnamefont {Schiller}}, \ and\ \bibinfo {author} {\bibfnamefont
  {S.}~\bibnamefont {Siem}},\ }\href {\doibase 10.1103/PhysRevLett.93.142504}
  {\bibfield  {journal} {\bibinfo  {journal} {Phys. Rev. Lett.}\ }\textbf
  {\bibinfo {volume} {93}},\ \bibinfo {pages} {142504} (\bibinfo {year}
  {2004})}\BibitemShut {NoStop}%
\bibitem [{\citenamefont {Larsen}\ \emph {et~al.}(2007)\citenamefont {Larsen},
  \citenamefont {Guttormsen}, \citenamefont {Chankova}, \citenamefont
  {Ingebretsen}, \citenamefont {L\"onnroth}, \citenamefont {Messelt},
  \citenamefont {Rekstad}, \citenamefont {Schiller}, \citenamefont {Siem},
  \citenamefont {Syed},\ and\ \citenamefont {Voinov}}]{Lar07}%
  \BibitemOpen
  \bibfield  {author} {\bibinfo {author} {\bibfnamefont {A.~C.}\ \bibnamefont
  {Larsen}}, \bibinfo {author} {\bibfnamefont {M.}~\bibnamefont {Guttormsen}},
  \bibinfo {author} {\bibfnamefont {R.}~\bibnamefont {Chankova}}, \bibinfo
  {author} {\bibfnamefont {F.}~\bibnamefont {Ingebretsen}}, \bibinfo {author}
  {\bibfnamefont {T.}~\bibnamefont {L\"onnroth}}, \bibinfo {author}
  {\bibfnamefont {S.}~\bibnamefont {Messelt}}, \bibinfo {author} {\bibfnamefont
  {J.}~\bibnamefont {Rekstad}}, \bibinfo {author} {\bibfnamefont
  {A.}~\bibnamefont {Schiller}}, \bibinfo {author} {\bibfnamefont
  {S.}~\bibnamefont {Siem}}, \bibinfo {author} {\bibfnamefont {N.~U.~H.}\
  \bibnamefont {Syed}}, \ and\ \bibinfo {author} {\bibfnamefont
  {A.}~\bibnamefont {Voinov}},\ }\href {\doibase 10.1103/PhysRevC.76.044303}
  {\bibfield  {journal} {\bibinfo  {journal} {Phys. Rev. C}\ }\textbf {\bibinfo
  {volume} {76}},\ \bibinfo {pages} {044303} (\bibinfo {year}
  {2007})}\BibitemShut {NoStop}%
\bibitem [{\citenamefont {Larsen}\ \emph
  {et~al.}(2013{\natexlab{a}})\citenamefont {Larsen}, \citenamefont {Ruud},
  \citenamefont {B\"urger}, \citenamefont {Goriely}, \citenamefont
  {Guttormsen}, \citenamefont {G\"orgen}, \citenamefont {Hagen}, \citenamefont
  {Harissopulos}, \citenamefont {Nyhus}, \citenamefont {Renstr\o{}m},
  \citenamefont {Schiller}, \citenamefont {Siem}, \citenamefont {Tveten},
  \citenamefont {Voinov},\ and\ \citenamefont {Wiedeking}}]{Lar13}%
  \BibitemOpen
  \bibfield  {author} {\bibinfo {author} {\bibfnamefont {A.~C.}\ \bibnamefont
  {Larsen}}, \bibinfo {author} {\bibfnamefont {I.~E.}\ \bibnamefont {Ruud}},
  \bibinfo {author} {\bibfnamefont {A.}~\bibnamefont {B\"urger}}, \bibinfo
  {author} {\bibfnamefont {S.}~\bibnamefont {Goriely}}, \bibinfo {author}
  {\bibfnamefont {M.}~\bibnamefont {Guttormsen}}, \bibinfo {author}
  {\bibfnamefont {A.}~\bibnamefont {G\"orgen}}, \bibinfo {author}
  {\bibfnamefont {T.~W.}\ \bibnamefont {Hagen}}, \bibinfo {author}
  {\bibfnamefont {S.}~\bibnamefont {Harissopulos}}, \bibinfo {author}
  {\bibfnamefont {H.~T.}\ \bibnamefont {Nyhus}}, \bibinfo {author}
  {\bibfnamefont {T.}~\bibnamefont {Renstr\o{}m}}, \bibinfo {author}
  {\bibfnamefont {A.}~\bibnamefont {Schiller}}, \bibinfo {author}
  {\bibfnamefont {S.}~\bibnamefont {Siem}}, \bibinfo {author} {\bibfnamefont
  {G.~M.}\ \bibnamefont {Tveten}}, \bibinfo {author} {\bibfnamefont
  {A.}~\bibnamefont {Voinov}}, \ and\ \bibinfo {author} {\bibfnamefont
  {M.}~\bibnamefont {Wiedeking}},\ }\href {\doibase 10.1103/PhysRevC.87.014319}
  {\bibfield  {journal} {\bibinfo  {journal} {Phys. Rev. C}\ }\textbf {\bibinfo
  {volume} {87}},\ \bibinfo {pages} {014319} (\bibinfo {year}
  {2013}{\natexlab{a}})}\BibitemShut {NoStop}%
\bibitem [{\citenamefont {Guttormsen}\ \emph {et~al.}(1987)\citenamefont
  {Guttormsen}, \citenamefont {Ramsøy},\ and\ \citenamefont
  {Rekstad}}]{GUTTORMSEN1987518}%
  \BibitemOpen
  \bibfield  {author} {\bibinfo {author} {\bibfnamefont {M.}~\bibnamefont
  {Guttormsen}}, \bibinfo {author} {\bibfnamefont {T.}~\bibnamefont {Ramsøy}},
  \ and\ \bibinfo {author} {\bibfnamefont {J.}~\bibnamefont {Rekstad}},\ }\href
  {\doibase https://doi.org/10.1016/0168-9002(87)91221-6} {\bibfield  {journal}
  {\bibinfo  {journal} {Nuclear Instruments and Methods in Physics Research
  Section A: Accelerators, Spectrometers, Detectors and Associated Equipment}\
  }\textbf {\bibinfo {volume} {255}},\ \bibinfo {pages} {518 } (\bibinfo {year}
  {1987})}\BibitemShut {NoStop}%
\bibitem [{\citenamefont {Guttormsen}\ \emph {et~al.}(1996)\citenamefont
  {Guttormsen}, \citenamefont {Tveter}, \citenamefont {Bergholt}, \citenamefont
  {Ingebretsen},\ and\ \citenamefont {Rekstad}}]{GUTTORMSEN1996371}%
  \BibitemOpen
  \bibfield  {author} {\bibinfo {author} {\bibfnamefont {M.}~\bibnamefont
  {Guttormsen}}, \bibinfo {author} {\bibfnamefont {T.}~\bibnamefont {Tveter}},
  \bibinfo {author} {\bibfnamefont {L.}~\bibnamefont {Bergholt}}, \bibinfo
  {author} {\bibfnamefont {F.}~\bibnamefont {Ingebretsen}}, \ and\ \bibinfo
  {author} {\bibfnamefont {J.}~\bibnamefont {Rekstad}},\ }\href {\doibase
  https://doi.org/10.1016/0168-9002(96)00197-0} {\bibfield  {journal} {\bibinfo
   {journal} {Nuclear Instruments and Methods in Physics Research Section A:
  Accelerators, Spectrometers, Detectors and Associated Equipment}\ }\textbf
  {\bibinfo {volume} {374}},\ \bibinfo {pages} {371 } (\bibinfo {year}
  {1996})}\BibitemShut {NoStop}%
\bibitem [{\citenamefont {Simon}\ \emph {et~al.}(2016)\citenamefont {Simon},
  \citenamefont {Guttormsen}, \citenamefont {Larsen}, \citenamefont {Beausang},
  \citenamefont {Humby}, \citenamefont {Burke}, \citenamefont {Casperson},
  \citenamefont {Hughes}, \citenamefont {Ross}, \citenamefont {Allmond},
  \citenamefont {Chyzh}, \citenamefont {Dag}, \citenamefont {Koglin},
  \citenamefont {McCleskey}, \citenamefont {McCleskey}, \citenamefont {Ota},\
  and\ \citenamefont {Saastamoinen}}]{PhysRevC.93.034303}%
  \BibitemOpen
  \bibfield  {author} {\bibinfo {author} {\bibfnamefont {A.}~\bibnamefont
  {Simon}}, \bibinfo {author} {\bibfnamefont {M.}~\bibnamefont {Guttormsen}},
  \bibinfo {author} {\bibfnamefont {A.~C.}\ \bibnamefont {Larsen}}, \bibinfo
  {author} {\bibfnamefont {C.~W.}\ \bibnamefont {Beausang}}, \bibinfo {author}
  {\bibfnamefont {P.}~\bibnamefont {Humby}}, \bibinfo {author} {\bibfnamefont
  {J.~T.}\ \bibnamefont {Burke}}, \bibinfo {author} {\bibfnamefont {R.~J.}\
  \bibnamefont {Casperson}}, \bibinfo {author} {\bibfnamefont {R.~O.}\
  \bibnamefont {Hughes}}, \bibinfo {author} {\bibfnamefont {T.~J.}\
  \bibnamefont {Ross}}, \bibinfo {author} {\bibfnamefont {J.~M.}\ \bibnamefont
  {Allmond}}, \bibinfo {author} {\bibfnamefont {R.}~\bibnamefont {Chyzh}},
  \bibinfo {author} {\bibfnamefont {M.}~\bibnamefont {Dag}}, \bibinfo {author}
  {\bibfnamefont {J.}~\bibnamefont {Koglin}}, \bibinfo {author} {\bibfnamefont
  {E.}~\bibnamefont {McCleskey}}, \bibinfo {author} {\bibfnamefont
  {M.}~\bibnamefont {McCleskey}}, \bibinfo {author} {\bibfnamefont
  {S.}~\bibnamefont {Ota}}, \ and\ \bibinfo {author} {\bibfnamefont
  {A.}~\bibnamefont {Saastamoinen}},\ }\href {\doibase
  10.1103/PhysRevC.93.034303} {\bibfield  {journal} {\bibinfo  {journal} {Phys.
  Rev. C}\ }\textbf {\bibinfo {volume} {93}},\ \bibinfo {pages} {034303}
  (\bibinfo {year} {2016})}\BibitemShut {NoStop}%
\bibitem [{\citenamefont {Kheswa}\ \emph {et~al.}(2015)\citenamefont {Kheswa},
  \citenamefont {Wiedeking}, \citenamefont {Giacoppo}, \citenamefont {Goriely},
  \citenamefont {Guttormsen}, \citenamefont {Larsen}, \citenamefont {Garrote},
  \citenamefont {Eriksen}, \citenamefont {Görgen}, \citenamefont {Hagen},
  \citenamefont {Koehler}, \citenamefont {Klintefjord}, \citenamefont {Nyhus},
  \citenamefont {Papka}, \citenamefont {Renstrøm}, \citenamefont {Rose},
  \citenamefont {Sahin}, \citenamefont {Siem},\ and\ \citenamefont
  {Tornyi}}]{Khe15}%
  \BibitemOpen
  \bibfield  {author} {\bibinfo {author} {\bibfnamefont {B.}~\bibnamefont
  {Kheswa}}, \bibinfo {author} {\bibfnamefont {M.}~\bibnamefont {Wiedeking}},
  \bibinfo {author} {\bibfnamefont {F.}~\bibnamefont {Giacoppo}}, \bibinfo
  {author} {\bibfnamefont {S.}~\bibnamefont {Goriely}}, \bibinfo {author}
  {\bibfnamefont {M.}~\bibnamefont {Guttormsen}}, \bibinfo {author}
  {\bibfnamefont {A.}~\bibnamefont {Larsen}}, \bibinfo {author} {\bibfnamefont
  {F.~B.}\ \bibnamefont {Garrote}}, \bibinfo {author} {\bibfnamefont
  {T.}~\bibnamefont {Eriksen}}, \bibinfo {author} {\bibfnamefont
  {A.}~\bibnamefont {Görgen}}, \bibinfo {author} {\bibfnamefont
  {T.}~\bibnamefont {Hagen}}, \bibinfo {author} {\bibfnamefont
  {P.}~\bibnamefont {Koehler}}, \bibinfo {author} {\bibfnamefont
  {M.}~\bibnamefont {Klintefjord}}, \bibinfo {author} {\bibfnamefont
  {H.}~\bibnamefont {Nyhus}}, \bibinfo {author} {\bibfnamefont
  {P.}~\bibnamefont {Papka}}, \bibinfo {author} {\bibfnamefont
  {T.}~\bibnamefont {Renstrøm}}, \bibinfo {author} {\bibfnamefont
  {S.}~\bibnamefont {Rose}}, \bibinfo {author} {\bibfnamefont {E.}~\bibnamefont
  {Sahin}}, \bibinfo {author} {\bibfnamefont {S.}~\bibnamefont {Siem}}, \ and\
  \bibinfo {author} {\bibfnamefont {T.}~\bibnamefont {Tornyi}},\ }\href
  {\doibase http://dx.doi.org/10.1016/j.physletb.2015.03.065} {\bibfield
  {journal} {\bibinfo  {journal} {Physics Letters B}\ }\textbf {\bibinfo
  {volume} {744}},\ \bibinfo {pages} {268 } (\bibinfo {year}
  {2015})}\BibitemShut {NoStop}%
\bibitem [{\citenamefont {Wiedeking}\ \emph {et~al.}(2012)\citenamefont
  {Wiedeking}, \citenamefont {Bernstein}, \citenamefont
  {Krti\ifmmode~\check{c}\else \v{c}\fi{}ka}, \citenamefont {Bleuel},
  \citenamefont {Allmond}, \citenamefont {Basunia}, \citenamefont {Burke},
  \citenamefont {Fallon}, \citenamefont {Firestone}, \citenamefont {Goldblum},
  \citenamefont {Hatarik}, \citenamefont {Lake}, \citenamefont {Lee},
  \citenamefont {Lesher}, \citenamefont {Paschalis}, \citenamefont {Petri},
  \citenamefont {Phair},\ and\ \citenamefont
  {Scielzo}}]{PhysRevLett.108.162503}%
  \BibitemOpen
  \bibfield  {author} {\bibinfo {author} {\bibfnamefont {M.}~\bibnamefont
  {Wiedeking}}, \bibinfo {author} {\bibfnamefont {L.~A.}\ \bibnamefont
  {Bernstein}}, \bibinfo {author} {\bibfnamefont {M.}~\bibnamefont
  {Krti\ifmmode~\check{c}\else \v{c}\fi{}ka}}, \bibinfo {author} {\bibfnamefont
  {D.~L.}\ \bibnamefont {Bleuel}}, \bibinfo {author} {\bibfnamefont {J.~M.}\
  \bibnamefont {Allmond}}, \bibinfo {author} {\bibfnamefont {M.~S.}\
  \bibnamefont {Basunia}}, \bibinfo {author} {\bibfnamefont {J.~T.}\
  \bibnamefont {Burke}}, \bibinfo {author} {\bibfnamefont {P.}~\bibnamefont
  {Fallon}}, \bibinfo {author} {\bibfnamefont {R.~B.}\ \bibnamefont
  {Firestone}}, \bibinfo {author} {\bibfnamefont {B.~L.}\ \bibnamefont
  {Goldblum}}, \bibinfo {author} {\bibfnamefont {R.}~\bibnamefont {Hatarik}},
  \bibinfo {author} {\bibfnamefont {P.~T.}\ \bibnamefont {Lake}}, \bibinfo
  {author} {\bibfnamefont {I.-Y.}\ \bibnamefont {Lee}}, \bibinfo {author}
  {\bibfnamefont {S.~R.}\ \bibnamefont {Lesher}}, \bibinfo {author}
  {\bibfnamefont {S.}~\bibnamefont {Paschalis}}, \bibinfo {author}
  {\bibfnamefont {M.}~\bibnamefont {Petri}}, \bibinfo {author} {\bibfnamefont
  {L.}~\bibnamefont {Phair}}, \ and\ \bibinfo {author} {\bibfnamefont {N.~D.}\
  \bibnamefont {Scielzo}},\ }\href {\doibase 10.1103/PhysRevLett.108.162503}
  {\bibfield  {journal} {\bibinfo  {journal} {Phys. Rev. Lett.}\ }\textbf
  {\bibinfo {volume} {108}},\ \bibinfo {pages} {162503} (\bibinfo {year}
  {2012})}\BibitemShut {NoStop}%
\bibitem [{\citenamefont {Schwengner}\ \emph {et~al.}(2013)\citenamefont
  {Schwengner}, \citenamefont {Frauendorf},\ and\ \citenamefont
  {Larsen}}]{PhysRevLett.111.232504}%
  \BibitemOpen
  \bibfield  {author} {\bibinfo {author} {\bibfnamefont {R.}~\bibnamefont
  {Schwengner}}, \bibinfo {author} {\bibfnamefont {S.}~\bibnamefont
  {Frauendorf}}, \ and\ \bibinfo {author} {\bibfnamefont {A.~C.}\ \bibnamefont
  {Larsen}},\ }\href {\doibase 10.1103/PhysRevLett.111.232504} {\bibfield
  {journal} {\bibinfo  {journal} {Phys. Rev. Lett.}\ }\textbf {\bibinfo
  {volume} {111}},\ \bibinfo {pages} {232504} (\bibinfo {year}
  {2013})}\BibitemShut {NoStop}%
\bibitem [{\citenamefont {Brown}\ and\ \citenamefont
  {Larsen}(2014)}]{PhysRevLett.113.252502}%
  \BibitemOpen
  \bibfield  {author} {\bibinfo {author} {\bibfnamefont {B.~A.}\ \bibnamefont
  {Brown}}\ and\ \bibinfo {author} {\bibfnamefont {A.~C.}\ \bibnamefont
  {Larsen}},\ }\href {\doibase 10.1103/PhysRevLett.113.252502} {\bibfield
  {journal} {\bibinfo  {journal} {Phys. Rev. Lett.}\ }\textbf {\bibinfo
  {volume} {113}},\ \bibinfo {pages} {252502} (\bibinfo {year}
  {2014})}\BibitemShut {NoStop}%
\bibitem [{\citenamefont {Sieja}(2017)}]{Sie17}%
  \BibitemOpen
  \bibfield  {author} {\bibinfo {author} {\bibfnamefont {K.}~\bibnamefont
  {Sieja}},\ }\href {\doibase 10.1103/PhysRevLett.119.052502} {\bibfield
  {journal} {\bibinfo  {journal} {Phys. Rev. Lett.}\ }\textbf {\bibinfo
  {volume} {119}},\ \bibinfo {pages} {052502} (\bibinfo {year}
  {2017})}\BibitemShut {NoStop}%
\bibitem [{\citenamefont {Litvinova}\ and\ \citenamefont
  {Belov}(2013)}]{PhysRevC.88.031302}%
  \BibitemOpen
  \bibfield  {author} {\bibinfo {author} {\bibfnamefont {E.}~\bibnamefont
  {Litvinova}}\ and\ \bibinfo {author} {\bibfnamefont {N.}~\bibnamefont
  {Belov}},\ }\href {\doibase 10.1103/PhysRevC.88.031302} {\bibfield  {journal}
  {\bibinfo  {journal} {Phys. Rev. C}\ }\textbf {\bibinfo {volume} {88}},\
  \bibinfo {pages} {031302} (\bibinfo {year} {2013})}\BibitemShut {NoStop}%
\bibitem [{\citenamefont {Larsen}\ \emph
  {et~al.}(2013{\natexlab{b}})\citenamefont {Larsen}, \citenamefont {Blasi},
  \citenamefont {Bracco}, \citenamefont {Camera}, \citenamefont {Eriksen},
  \citenamefont {G\"orgen}, \citenamefont {Guttormsen}, \citenamefont {Hagen},
  \citenamefont {Leoni}, \citenamefont {Million}, \citenamefont {Nyhus},
  \citenamefont {Renstr\o{}m}, \citenamefont {Rose}, \citenamefont {Ruud},
  \citenamefont {Siem}, \citenamefont {Tornyi}, \citenamefont {Tveten},
  \citenamefont {Voinov},\ and\ \citenamefont
  {Wiedeking}}]{PhysRevLett.111.242504}%
  \BibitemOpen
  \bibfield  {author} {\bibinfo {author} {\bibfnamefont {A.~C.}\ \bibnamefont
  {Larsen}}, \bibinfo {author} {\bibfnamefont {N.}~\bibnamefont {Blasi}},
  \bibinfo {author} {\bibfnamefont {A.}~\bibnamefont {Bracco}}, \bibinfo
  {author} {\bibfnamefont {F.}~\bibnamefont {Camera}}, \bibinfo {author}
  {\bibfnamefont {T.~K.}\ \bibnamefont {Eriksen}}, \bibinfo {author}
  {\bibfnamefont {A.}~\bibnamefont {G\"orgen}}, \bibinfo {author}
  {\bibfnamefont {M.}~\bibnamefont {Guttormsen}}, \bibinfo {author}
  {\bibfnamefont {T.~W.}\ \bibnamefont {Hagen}}, \bibinfo {author}
  {\bibfnamefont {S.}~\bibnamefont {Leoni}}, \bibinfo {author} {\bibfnamefont
  {B.}~\bibnamefont {Million}}, \bibinfo {author} {\bibfnamefont {H.~T.}\
  \bibnamefont {Nyhus}}, \bibinfo {author} {\bibfnamefont {T.}~\bibnamefont
  {Renstr\o{}m}}, \bibinfo {author} {\bibfnamefont {S.~J.}\ \bibnamefont
  {Rose}}, \bibinfo {author} {\bibfnamefont {I.~E.}\ \bibnamefont {Ruud}},
  \bibinfo {author} {\bibfnamefont {S.}~\bibnamefont {Siem}}, \bibinfo {author}
  {\bibfnamefont {T.}~\bibnamefont {Tornyi}}, \bibinfo {author} {\bibfnamefont
  {G.~M.}\ \bibnamefont {Tveten}}, \bibinfo {author} {\bibfnamefont {A.~V.}\
  \bibnamefont {Voinov}}, \ and\ \bibinfo {author} {\bibfnamefont
  {M.}~\bibnamefont {Wiedeking}},\ }\href {\doibase
  10.1103/PhysRevLett.111.242504} {\bibfield  {journal} {\bibinfo  {journal}
  {Phys. Rev. Lett.}\ }\textbf {\bibinfo {volume} {111}},\ \bibinfo {pages}
  {242504} (\bibinfo {year} {2013}{\natexlab{b}})}\BibitemShut {NoStop}%
\bibitem [{\citenamefont {Larsen}\ \emph {et~al.}(2017)\citenamefont {Larsen},
  \citenamefont {Guttormsen}, \citenamefont {Blasi}, \citenamefont {Bracco},
  \citenamefont {Camera}, \citenamefont {Campo}, \citenamefont {Eriksen},
  \citenamefont {Görgen}, \citenamefont {Hagen}, \citenamefont {Ingeberg},
  \citenamefont {Kheswa}, \citenamefont {Leoni}, \citenamefont {Midtbø},
  \citenamefont {Million}, \citenamefont {Nyhus}, \citenamefont {Renstrøm},
  \citenamefont {Rose}, \citenamefont {Ruud}, \citenamefont {Siem},
  \citenamefont {Tornyi}, \citenamefont {Tveten}, \citenamefont {Voinov},
  \citenamefont {Wiedeking},\ and\ \citenamefont {Zeiser}}]{Lar17}%
  \BibitemOpen
  \bibfield  {author} {\bibinfo {author} {\bibfnamefont {A.~C.}\ \bibnamefont
  {Larsen}}, \bibinfo {author} {\bibfnamefont {M.}~\bibnamefont {Guttormsen}},
  \bibinfo {author} {\bibfnamefont {N.}~\bibnamefont {Blasi}}, \bibinfo
  {author} {\bibfnamefont {A.}~\bibnamefont {Bracco}}, \bibinfo {author}
  {\bibfnamefont {F.}~\bibnamefont {Camera}}, \bibinfo {author} {\bibfnamefont
  {L.~C.}\ \bibnamefont {Campo}}, \bibinfo {author} {\bibfnamefont {T.~K.}\
  \bibnamefont {Eriksen}}, \bibinfo {author} {\bibfnamefont {A.}~\bibnamefont
  {Görgen}}, \bibinfo {author} {\bibfnamefont {T.~W.}\ \bibnamefont {Hagen}},
  \bibinfo {author} {\bibfnamefont {V.~W.}\ \bibnamefont {Ingeberg}}, \bibinfo
  {author} {\bibfnamefont {B.~V.}\ \bibnamefont {Kheswa}}, \bibinfo {author}
  {\bibfnamefont {S.}~\bibnamefont {Leoni}}, \bibinfo {author} {\bibfnamefont
  {J.~E.}\ \bibnamefont {Midtbø}}, \bibinfo {author} {\bibfnamefont
  {B.}~\bibnamefont {Million}}, \bibinfo {author} {\bibfnamefont {H.~T.}\
  \bibnamefont {Nyhus}}, \bibinfo {author} {\bibfnamefont {T.}~\bibnamefont
  {Renstrøm}}, \bibinfo {author} {\bibfnamefont {S.~J.}\ \bibnamefont {Rose}},
  \bibinfo {author} {\bibfnamefont {I.~E.}\ \bibnamefont {Ruud}}, \bibinfo
  {author} {\bibfnamefont {S.}~\bibnamefont {Siem}}, \bibinfo {author}
  {\bibfnamefont {T.~G.}\ \bibnamefont {Tornyi}}, \bibinfo {author}
  {\bibfnamefont {G.~M.}\ \bibnamefont {Tveten}}, \bibinfo {author}
  {\bibfnamefont {A.~V.}\ \bibnamefont {Voinov}}, \bibinfo {author}
  {\bibfnamefont {M.}~\bibnamefont {Wiedeking}}, \ and\ \bibinfo {author}
  {\bibfnamefont {F.}~\bibnamefont {Zeiser}},\ }\href
  {http://stacks.iop.org/0954-3899/44/i=6/a=064005} {\bibfield  {journal}
  {\bibinfo  {journal} {Journal of Physics G: Nuclear and Particle Physics}\
  }\textbf {\bibinfo {volume} {44}},\ \bibinfo {pages} {064005} (\bibinfo
  {year} {2017})}\BibitemShut {NoStop}%
\bibitem [{\citenamefont {Voinov}\ \emph {et~al.}(2010)\citenamefont {Voinov},
  \citenamefont {Grimes}, \citenamefont {Brune}, \citenamefont {Guttormsen},
  \citenamefont {Larsen}, \citenamefont {Massey}, \citenamefont {Schiller},\
  and\ \citenamefont {Siem}}]{PhysRevC.81.024319}%
  \BibitemOpen
  \bibfield  {author} {\bibinfo {author} {\bibfnamefont {A.}~\bibnamefont
  {Voinov}}, \bibinfo {author} {\bibfnamefont {S.~M.}\ \bibnamefont {Grimes}},
  \bibinfo {author} {\bibfnamefont {C.~R.}\ \bibnamefont {Brune}}, \bibinfo
  {author} {\bibfnamefont {M.}~\bibnamefont {Guttormsen}}, \bibinfo {author}
  {\bibfnamefont {A.~C.}\ \bibnamefont {Larsen}}, \bibinfo {author}
  {\bibfnamefont {T.~N.}\ \bibnamefont {Massey}}, \bibinfo {author}
  {\bibfnamefont {A.}~\bibnamefont {Schiller}}, \ and\ \bibinfo {author}
  {\bibfnamefont {S.}~\bibnamefont {Siem}},\ }\href {\doibase
  10.1103/PhysRevC.81.024319} {\bibfield  {journal} {\bibinfo  {journal} {Phys.
  Rev. C}\ }\textbf {\bibinfo {volume} {81}},\ \bibinfo {pages} {024319}
  (\bibinfo {year} {2010})}\BibitemShut {NoStop}%
\bibitem [{\citenamefont {Voinov}\ and\ \citenamefont
  {Grimes}(2015)}]{PhysRevC.92.064308}%
  \BibitemOpen
  \bibfield  {author} {\bibinfo {author} {\bibfnamefont {A.~V.}\ \bibnamefont
  {Voinov}}\ and\ \bibinfo {author} {\bibfnamefont {S.~M.}\ \bibnamefont
  {Grimes}},\ }\href {\doibase 10.1103/PhysRevC.92.064308} {\bibfield
  {journal} {\bibinfo  {journal} {Phys. Rev. C}\ }\textbf {\bibinfo {volume}
  {92}},\ \bibinfo {pages} {064308} (\bibinfo {year} {2015})}\BibitemShut
  {NoStop}%
\bibitem [{\citenamefont {Paschalis}\ \emph {et~al.}(2013)\citenamefont
  {Paschalis} \emph {et~al.}}]{PASCHALIS201344}%
  \BibitemOpen
  \bibfield  {author} {\bibinfo {author} {\bibfnamefont {S.}~\bibnamefont
  {Paschalis}} \emph {et~al.},\ }\href {\doibase
  http://dx.doi.org/10.1016/j.nima.2013.01.009} {\bibfield  {journal} {\bibinfo
   {journal} {Nucl. Instrum. Methods Phys. Res., Sect. A}\ }\textbf {\bibinfo
  {volume} {709}},\ \bibinfo {pages} {44 } (\bibinfo {year}
  {2013})}\BibitemShut {NoStop}%
\bibitem [{\citenamefont {Sarantites}\ \emph {et~al.}(2015)\citenamefont
  {Sarantites}, \citenamefont {Reviol} \emph {et~al.}}]{Sarantites201542}%
  \BibitemOpen
  \bibfield  {author} {\bibinfo {author} {\bibfnamefont {D.}~\bibnamefont
  {Sarantites}}, \bibinfo {author} {\bibfnamefont {W.}~\bibnamefont {Reviol}},
  \emph {et~al.},\ }\href {\doibase https://doi.org/10.1016/j.nima.2015.04.007}
  {\bibfield  {journal} {\bibinfo  {journal} {Nucl. Instrum. Methods Phys.
  Res., Sect. A}\ }\textbf {\bibinfo {volume} {790}},\ \bibinfo {pages} {42 }
  (\bibinfo {year} {2015})}\BibitemShut {NoStop}%
\bibitem [{\citenamefont {Junde}\ \emph {et~al.}(2011)\citenamefont {Junde},
  \citenamefont {Su},\ and\ \citenamefont {Dong}}]{JUNDE20111513}%
  \BibitemOpen
  \bibfield  {author} {\bibinfo {author} {\bibfnamefont {H.}~\bibnamefont
  {Junde}}, \bibinfo {author} {\bibfnamefont {H.}~\bibnamefont {Su}}, \ and\
  \bibinfo {author} {\bibfnamefont {Y.}~\bibnamefont {Dong}},\ }\href {\doibase
  http://dx.doi.org/10.1016/j.nds.2011.04.004} {\bibfield  {journal} {\bibinfo
  {journal} {Nuclear Data Sheets}\ }\textbf {\bibinfo {volume} {112}},\
  \bibinfo {pages} {1513 } (\bibinfo {year} {2011})}\BibitemShut {NoStop}%
\bibitem [{\citenamefont {Mateosian}\ and\ \citenamefont
  {Sunyar}(1974)}]{DERMATEOSIAN1974391}%
  \BibitemOpen
  \bibfield  {author} {\bibinfo {author} {\bibfnamefont {E.~D.}\ \bibnamefont
  {Mateosian}}\ and\ \bibinfo {author} {\bibfnamefont {A.}~\bibnamefont
  {Sunyar}},\ }\href {\doibase http://dx.doi.org/10.1016/0092-640X(74)90007-2}
  {\bibfield  {journal} {\bibinfo  {journal} {Atomic Data and Nuclear Data
  Tables}\ }\textbf {\bibinfo {volume} {13}},\ \bibinfo {pages} {391 }
  (\bibinfo {year} {1974})}\BibitemShut {NoStop}%
\bibitem [{\citenamefont {Alikhani}\ \emph {et~al.}(2012)\citenamefont
  {Alikhani}, \citenamefont {Givechev}, \citenamefont {Heinz}, \citenamefont
  {John}, \citenamefont {Leske}, \citenamefont {Lettmann}, \citenamefont
  {Möller}, \citenamefont {Pietralla},\ and\ \citenamefont
  {Röder}}]{ALIKHANI2012144}%
  \BibitemOpen
  \bibfield  {author} {\bibinfo {author} {\bibfnamefont {B.}~\bibnamefont
  {Alikhani}}, \bibinfo {author} {\bibfnamefont {A.}~\bibnamefont {Givechev}},
  \bibinfo {author} {\bibfnamefont {A.}~\bibnamefont {Heinz}}, \bibinfo
  {author} {\bibfnamefont {P.}~\bibnamefont {John}}, \bibinfo {author}
  {\bibfnamefont {J.}~\bibnamefont {Leske}}, \bibinfo {author} {\bibfnamefont
  {M.}~\bibnamefont {Lettmann}}, \bibinfo {author} {\bibfnamefont
  {O.}~\bibnamefont {Möller}}, \bibinfo {author} {\bibfnamefont
  {N.}~\bibnamefont {Pietralla}}, \ and\ \bibinfo {author} {\bibfnamefont
  {C.}~\bibnamefont {Röder}},\ }\href {\doibase
  http://dx.doi.org/10.1016/j.nima.2012.02.016} {\bibfield  {journal} {\bibinfo
   {journal} {Nuclear Instruments and Methods in Physics Research Section A:
  Accelerators, Spectrometers, Detectors and Associated Equipment}\ }\textbf
  {\bibinfo {volume} {675}},\ \bibinfo {pages} {144 } (\bibinfo {year}
  {2012})}\BibitemShut {NoStop}%
\bibitem [{\citenamefont {Wiens}(2014)}]{AWiens}%
  \BibitemOpen
  \bibfield  {author} {\bibinfo {author} {\bibfnamefont {A.}~\bibnamefont
  {Wiens}},\ }\href {http://meetings.aps.org/link/BAPS.2014.HAW.DK.2}
  {\bibfield  {journal} {\bibinfo  {journal} {4$^{th}$ Joint Meeting of the APS
  Division of Nuclear Physics and Physical Society of Japan}\ } (\bibinfo
  {year} {2014})}\BibitemShut {NoStop}%
\bibitem [{\citenamefont {Fagg}\ and\ \citenamefont
  {Hanna}(1959)}]{RevModPhys.31.711}%
  \BibitemOpen
  \bibfield  {author} {\bibinfo {author} {\bibfnamefont {L.~W.}\ \bibnamefont
  {Fagg}}\ and\ \bibinfo {author} {\bibfnamefont {S.~S.}\ \bibnamefont
  {Hanna}},\ }\href {\doibase 10.1103/RevModPhys.31.711} {\bibfield  {journal}
  {\bibinfo  {journal} {Rev. Mod. Phys.}\ }\textbf {\bibinfo {volume} {31}},\
  \bibinfo {pages} {711} (\bibinfo {year} {1959})}\BibitemShut {NoStop}%
\end{thebibliography}%

\end{document}